\documentclass[12pt]{aastex61}
\usepackage{amsmath,amssymb}
\usepackage{float}
\usepackage{natbib}    
\usepackage{color}
\usepackage{hyperref} 

\def\deg{\ifmmode^\circ\else$^\circ$\fi}
\def\degc{\ifmmode^\circ\hbox{C}\else$^\circ$C\fi}
\def\arcsec{\ifmmode^{\prime\prime}\else$^{\prime\prime}$\fi}
\def\arcmin{\ifmmode^{\prime}\else$^{\prime}$\fi}

\def\earth{\ifmmode_{\mathord\oplus}\else$_{\mathord\oplus}$\fi}

\shorttitle{ATLAS Transient Survey}
\shortauthors{Smith et al.}

\begin{document}

\title{Design and operation of the ATLAS Transient Science Server}

\author{K. W. Smith}
\affiliation{Astrophysics Research Centre, School of Mathematics and Physics, Queen's University Belfast, Belfast, BT7 1NN, UK}
\author{S. J. Smartt}
\affiliation{Astrophysics Research Centre, School of Mathematics and Physics, Queen's University Belfast, Belfast, BT7 1NN, UK}
\author{D. R. Young}
\affiliation{Astrophysics Research Centre, School of Mathematics and Physics, Queen's University Belfast, Belfast, BT7 1NN, UK}
\author{J. L. Tonry}
\affiliation{Institute for Astronomy, University of Hawaii, 2680 Woodlawn Drive, Honolulu, HI 96822}
\author{L. Denneau}
\affiliation{Institute for Astronomy, University of Hawaii, 2680 Woodlawn Drive, Honolulu, HI 96822}
\author{H. Flewelling}
\affiliation{Institute for Astronomy, University of Hawaii, 2680 Woodlawn Drive, Honolulu, HI 96822}
\author{A. N. Heinze}
\affiliation{Institute for Astronomy, University of Hawaii, 2680 Woodlawn Drive, Honolulu, HI 96822}
\author{H. J. Weiland}
\affiliation{Institute for Astronomy, University of Hawaii, 2680 Woodlawn Drive, Honolulu, HI 96822}
\author{B. Stalder}
\affiliation{Vera C. Rubin Observatory Project Office, 950 N Cherry Ave, Tucson, AZ 85719, USA}
\author{A. Rest}
\affiliation{Space Telescope Science Institute, 3700 San Martin Drive, Baltimore, MD 21218, USA}
\affiliation{Department of Physics and Astronomy, Johns Hopkins University, Baltimore, MD 21218, USA}
\author{C. W. Stubbs}
\affiliation{Department of Physics, Harvard University, Cambridge, MA 02138, USA}
\author{J. P. Anderson}
\affiliation{European Southern Observatory, Alonso de Córdova 3107, Casilla 19 Santiago, Chile}
\affiliation{Millennium Institute of Astrophysics, Chile}
\author{T.-W Chen}
\affiliation{The Oskar Klein Centre, Department of Astronomy, Stockholm University, AlbaNova, SE-10691 Stockholm, Sweden}
\author{P. Clark}
\affiliation{Astrophysics Research Centre, School of Mathematics and Physics, Queen's University Belfast, Belfast, BT7 1NN, UK}
\author{A. Do}
\affiliation{Institute for Astronomy, University of Hawaii, 2680 Woodlawn Drive, Honolulu, HI 96822}
\author{F. F\"{o}rster}
\affiliation{Millennium Institute of Astrophysics, Chile}
\affiliation{Center for Mathematical Modeling, Universidad de Chile, Chile}
\author{M. Fulton}
\affiliation{Astrophysics Research Centre, School of Mathematics and Physics, Queen's University Belfast, Belfast, BT7 1NN, UK}
\author{J. Gillanders}
\affiliation{Astrophysics Research Centre, School of Mathematics and Physics, Queen's University Belfast, Belfast, BT7 1NN, UK}
\author{O. R. McBrien}
\affiliation{Astrophysics Research Centre, School of Mathematics and Physics, Queen's University Belfast, Belfast, BT7 1NN, UK}
\author{D. O'Neill}
\affiliation{Astrophysics Research Centre, School of Mathematics and Physics, Queen's University Belfast, Belfast, BT7 1NN, UK}
\author{S. Srivastav}
\affiliation{Astrophysics Research Centre, School of Mathematics and Physics, Queen's University Belfast, Belfast, BT7 1NN, UK}
\author{D. E. Wright}
\affiliation{School of Physics and Astronomy, 116 Church Street SE, University of Minnesota, Minneapolis, MN 55455, USA}

\begin{abstract}
The  Asteroid Terrestrial impact Last Alert System (ATLAS) system  consists of two 0.5m Schmidt telescopes  with cameras covering 
29 square degrees at plate scale of 1.86 arcsec per pixel. Working in tandem, the telescopes  routinely survey the whole sky visible from Hawaii (above $\delta > -50^{\circ}$) every two nights, exposing four times per night, typically reaching $o < 19$ magnitude \emph{per exposure} when the moon is illuminated and $c < 19.5$ magnitude \emph{per exposure} in dark skies. Construction is underway of two further units to be sited in Chile and South Africa which will result in an all-sky daily cadence from 2021.
Initially designed for detecting potentially hazardous near earth objects, the ATLAS data enable a range of astrophysical time domain science. To extract transients from the data stream requires a computing system to process the data, assimilate detections in time and space and associate them with known astrophysical sources. Here we describe the hardware and software infrastructure to produce a stream of clean, real, astrophysical transients in real time. This involves machine learning and boosted decision tree algorithms to identify extragalactic and Galactic transients. Typically we detect 10-15 supernova candidates per night which we immediately announce publicly.  
 The ATLAS discoveries not only enable rapid follow-up of interesting  sources but will provide complete statistical samples within the local volume of 100\,Mpc. A simple comparison of the detected supernova rate within 100\,Mpc, with no corrections for completeness, is already significantly higher (factor 1.5 to 2) than the current accepted rates.

\end{abstract}



\section{Introduction}
\label{sec:intro}

The use of ground-based wide-field telescopes equipped with large format digital cameras has 
opened a window on the time varying Universe over the last decade. The late 1990s and first 
decade of the twentieth century saw relatively high redshift surveys for type Ia supernovae
\cite[e.g.][]{1998ApJ...507...46S,2006A&A...447...31A}
and low redshift surveys focused on galaxy targeted searches
\citep{2011MNRAS.412.1419L,2011MNRAS.412.1441L}
The former typically 
employed what were then regarded as wide-field CCD camera mosaics on 2-4m telescopes
\citep[of order 0.5$^{\circ}$ to 1$^{\circ}$ diameter field-of-view;][]{1997ApJ...483..565P}
with carefully selected 
targets for spectroscopic follow-up on the new and precious
8-10 metre aperture facilities
\citep{2005ApJ...634.1190H,2007ApJ...666..694W}.

Low redshift
supernovae (SNe) were found primarily by narrow field, small telescopes 
\cite[0.5 - 1. metre;][]{2009AIPC.1111..551P,2011MNRAS.412.1419L} and by 
well equipped and experienced amateur astronomers. 
These bright events were subject to follow-up by many teams 
around the globe with moderate sized telescopes and many were observed and detected from x-rays
\citep{2003LNP...598...91I,2006ApJ...648L.119I,2008Natur.453..469S}
to ultraviolet \citep{2009AJ....137.4517B,2009ApJ...700.1456B}  and from the mid-infrared 
\citep{2006ApJ...651L.117K,2011MNRAS.418.1285F}
to radio \citep{2003A&A...397.1011S,2004MNRAS.349.1093R,2007ApJ...671..689S,2010Natur.463..513S,2017ApJ...835..140M}. 
Both the low and high redshift surveys were highly 
successful in their respective goals and in particular in allowing understanding of the 
physics of type Ia SNe to shape their use in cosmology.  

The successful techniques of both these approaches have been combined in the last decade to 
allow wide-field, virtually all sky, surveys of the local Universe with 10 to 100 sq degree
cameras \citep{Quimby2006PhD,2009PASP..121.1395L,2009ApJ...696..870D,2016arXiv161205560C,2013PASP..125..683B,2017MNRAS.464.2672H,2018PASP..130f4505T,2019PASP..131a8002B}. 
This allows a large volume to be searched and importantly this volume samples
the local Universe, enabling extensive multi-wavelength monitoring of bright objects.
Photon starvation from distant objects tends to raise barriers to physical understanding
and development of radiative transfer simulations, shock physics models and theories of 
powering mechanisms \cite[e.g.][]{2009MNRAS.398.1809K,2010ApJ...717..245K,2011ApJ...728...63R}
The rate of discovery of supernovae and transients within $z \lesssim 0.1$
increased dramatically and significant numbers of 
transients were found either in faint host galaxies or in environments far 
from their high mass hosts (tens of kiloparsecs). 
The combination of the sheer number of discoveries and  different 
explosion environments probed in comparison to the historic surveys revealed
an unexpected and remarkable diversity in types of stellar explosions and merger events 
occurring in the local Universe. An important scientific advance was combining the 
discovery of transients with rapid spectroscopic follow-up to determine redshifts, 
association with host galaxies and chemical composition of the ejecta 
\cite[e.g.][]{2011Natur.474..487Q,2013ApJ...770..128I,2014Natur.509..471G}. 
Furthermore, as these 
low redshift transients were relatively bright then multi-wavelength, time domain follow-up has 
unveiled new explosion physics. And remarkably for the first time, a particle 
messenger other than a photon provided an alternative alert that triggered exact 
identification of the source with optical-near infrared telescopes: the joint discovery of 
GW170817, GRB170817A and AT2017gfo \citep{2017PhRvL.119p1101A,2017ApJ...848L..13A,2017ApJ...848L..12A}\footnote{While multi-messenger astrophysics did begin with solar and supernova neutrinos, neither of these photon emitting objects were discovered due to the directional information of their neutrino detections. AT2017gfo was discovered by astronomers directly following an alert from a non-photonic detector.}.

A critical piece of a wide-field imaging survey and discovery facility is the computing 
software and hardware that processes the data to produce a stream of clean, real, astrophysical 
variables and transients. This comprises two parts, the first being the detector detrending, 
source identification and astrometric and photometric calibration
\citep[e.g.][]{waters2017}.
The principles of such processing are based on well known 
methods applied to narrow-field telescopes, although wide-field camera calibration has 
its own challenges
\citep{magnier2017a,magnier2017b}.
The now ubiquitous use of difference imaging has provided a method to
identify and quantify variables and transient sources 
\citep{1998ApJ...503..325A,2015ascl.soft04004B,2016ApJ...830...27Z}.
However at that point the night's catalogues of objects
still requires further processing to provide a useful information stream that can be used 
scientifically, and immediately, by astronomers. 
This means filtering 
the bogus detections that inevitably 
exist on the difference images of a CCD, associating the detections together and 
assimilating them into distinct objects and providing basic classifications
\citep{2013MNRAS.435.1047B,2015MNRAS.449..451W,2015AJ....150...82G,2019PASP..131c8002M,2018ApJS..236....9N}.
Real astrophysical detections on a difference image comprise satellite trails, minor planets, 
fast moving near earth asteroids (point sources and trails), variable and erupting stars in the
Milky Way and Local Group galaxies, novae, active galactic nuclei (AGN), tidal disruption events
and extraglactic supernovae, stellar mergers and afterglows of gamma-ray-bursts. 
Typically
most wide-field survey and discovery projects have built a complex software system that 
provides a means of extracting and classifying the sources and feeding them to spectroscopic follow-up 
programmes \citep{2015A&A...579A..40S,2018PASP..130c5003B,2019PASP..131g8001G}. Classification of the
lightcurves to filter and select objects 
of interest has had some success 
\citep{2016ApJS..225...31L,2019PASP..131k8002M}. 

We will use the term ``transient science server'' for this system of software and computing hardware. The back end of these systems which focused on the assimilation of individual detections into object lightcurves and classification of the objects from their host object (star or galaxy) have recently been termed ``brokers''. The Zwicky Transient Facility 
\citep{2019PASP..131a8002B} 
issues a public alert stream of transient and variable 
sources detected on their difference images
\citep{2019PASP..131a8001P}, 
with a 
view to testing methods for the upcoming 
Rubin Observatory Legacy Survey of Space and Time (LSST). 
These alert packets are ingested and processed by a 
series of ``brokers'' to provide added value 
such as context (variable star and AGN identification, host galaxy, redshift, association 
with known transients, variables or outbursting source) and lightcurve classification. Currently four systems 
ingest these data and provide different functionality 
to users, Lasair\footnote{\url{https://lasair.roe.ac.uk/}} \citep{2019RNAAS...3...26S}
ANTARES\footnote{\url{https://antares.noao.edu/}} \citep{2018ApJS..236....9N}, 
ALeRCE\footnote{\url{https://alerce.online/}} \citep[][white paper in prep.]{2019TNSTR1403....1F}
and MARS\footnote{\url{https://mars.lco.global/}}. 

In this paper we describe how the ATLAS  Transient Science Server functions and operates on a daily basis. The ATLAS system was defined in \cite{2018PASP..130f4505T} and the first major data release (a variable star catalogue) was described in \cite{2018AJ....156..241H}.  ATLAS is designed, funded, and operated using NASA funds to find
 dangerous asteroids which are routinely detected and submitted to the
 Minor Planet Center (Denneau et al. in prep).  The nature of the search for moving objects
 yields detections of variable and transient objects for free. To date, around
25 science papers have been published on transient
and variable objects from the ATLAS data processing. 
This includes the discovery of AT2018cow
\citep{2018ApJ...865L...3P}
a candidate for a pulsational pair instability 
supernova 
\citep{2018ApJ...867L..31C}
a possible white dwarf - neutron star
merger
\citep{2019ApJ...885L..23M} and the lowest luminosity
supernova ever discovered
\citep{2020arXiv200109722S}. ATLAS has covered the gravitational wave skymaps of 
a number of LIGO Virgo events, often covering a complete fraction of the map and 
discovering candidates rapidly. For example we 
discovered a fast fading transient in the skymap of GW170104, which turned out 
to be the afterglow of a gamma ray burst (GRB170105A) but was discovered without the 
high energy trigger \citep{2017ApJ...850..149S}.  After describing the system in 
Sections\,\ref{sec:telsurvey} - \ref{sec:onsky}, we briefly outline the scientific return 
in Section\,\ref{sec:sci}
and  improvements for the future in Section\,\ref{sec:conc}.

\section{The ATLAS Telescopes and survey description}
\label{sec:telsurvey}
ATLAS currently consists of two identical telescopes which are a 
variant of a ``Wright Schmidt'' design. They consist of a 0.5~m Schmidt
corrector, a 0.65~m spherical primary mirror, and a three lens field 
corrector sitting in front of the filter changer and camera. 
They are situated at the Haleakala Observatory (HKO) on Maui and the Mauna Loa Observatory (MLO) on the Big Island of Hawaii. The first telescope (on HKO) has been fully operational since 2015, and the second (on MLO) was commissioned at the end of January 2017.  The telescopes use two main filters in normal survey operations. One is ``cyan'' (we refer to it as $c$) which is 
used during dark time and is roughly equivalent to a composite PS1 $g+r$
(420-650\,nm). The other is 
 ``orange'' (or $o$),  used during bright time  and  roughly equivalent to $r+i$ (560-820\,nm). The Haleakala unit hereafter referred to as ATLAS-HKO has both the $c$ and $o$
filters and the the Mauana Loa unit (ATLAS-MLO) has only the $o$ filter. 

The cameras on each telescope (which we call an ``ACAM'') each contain a STA 1600 $10560\times 10560$ pixel  single CCD detector.  The pixel scale is 1.86 arcsec per pixel, and excluding the borders of the CCDs (a 30 pixel boundary) the net field of view is $5.4^{\circ}\times5.4^{\circ}$, equating to 29 square degrees. 
In normal survey operations, the telescope schedules are jointly planned to 
tile the sky with 30 second exposures. As described in 
\cite{2018PASP..130f4505T}, 
the overhead from shutter closure to beginning of the next frame is 
10 sec. 
This is determined by the readtime of the chip (which is $\sim$9\,sec)
since the ATLAS telescope mounts are fast and reliable. As described in
\cite{2018PASP..130f4505T} 
for telescope movements smaller than 45$^{\circ}$ it takes 
$6.5\pm0.8$ sec to slew and resume tracking. 
Hence a total duration (expose plus overheads) of 
40\,sec per frame means that  
each telescope can take a \emph{maximum} of about 
1000 
exposures per night during the longest night duration of 
11.5\,hrs.  In practice, between 800 and 1000 frames  per telescope per
night are routinely taken (see Figure\,\ref{fig:Ingestion}), with 900 frames
covering 26\,100 square degrees. 
This means each of the telescopes could comfortably cover 
the entire visible sky from Hawaii ($\sim$24\,500 deg$^{2}$: north celestial pole down to about $-45$ degrees declination) every night. The entire funding for ATLAS comes from NASA to find dangerous asteroids; hence operations are designed to optimise the detectability of moving objects.  It is mandatory to collect 4 views of each field every night over a span of less than 1 hour in order to reach a manageable false alarm rate. 
This facilitates identification of moving objects, linkage of the tracklets,  orbit estimation 
and submission to the Minor Planet Center.  
Hence the total sky footprint covered by each telescope, 4 times each night, is around 6500 square degrees. The repeat field coverage also enables intra night (and multi night) difference stacks to be created, allowing ATLAS to detect slow moving and static transients at $m>20$ mag across the whole sky, including the galactic plane.

The telescopes are scheduled to  scan declination strips of the sky and, weather permitting, each will return to the same footprint
area every 4 nights.  Hence with two units working in tandem, the whole sky visible from Hawaii is covered every 2 nights with a 
$4 \times 30$\,sec observing sequence.  A typical 2 day footprint is 
illustrated in Figure\,5 of \cite{2018PASP..130f4505T}. 
The southern pointing limit
is a boresight of $\delta=-47.5^{\circ}$, and the southern egde of these
exposures reach $\delta=-50^{\circ}$. 
The survey strategy
has been defined to optimise for NEO cadence
\citep[e.g.][]{2009Icar..203..472V,2011PASP..123...58T,2018PASP..130f4505T}, 
but a 2-day, all sky cadence is an excellent survey base for 
early discovery and lightcurve classification of stationary transients. 

\section{ATLAS data processing and object detection}
The data processing steps to detrend, process and to photometrically and astrometrically 
analyse the images  are described in 
\cite{2018PASP..130f4505T}, 
along with all the timings for each step (see Table\, 3 in that paper).  
Each 30\,sec frame is processed on summit computers and it takes 310
seconds to apply the first initial step of 
detector detrending and photometric and astrometric
calibration. It requires a further 1280 seconds to produce a matching reference image (from the ATLAS wallpaper), subtract it from the 
input image, detect and classify the sources and write out a file containing
the detections with $S/N > 5\sigma$. Hence a total of 27 minutes
is required from the time of shutter opening for an exposure to 
the final processed file containing a list of detections. The 
processing runs multi-threaded, and not in serial, hence 27 minutes
is the typical delay time from observing to initially quantifying detections
on the cameras. We have the capability of reducing the elapsed time by factor of 
$\sim$2  through deployment of multi-threading on the summit computers, but currently 
do not do this since the primary science goal of asteroid detection has
a latency of $\sim$30\,mins due to the 4 visits.

A key part of transient object detection is subtraction of a 
reference sky frame from the input image to allow for rapid and 
automated detection of variables, moving objects and transients
\citep{1996AJ....112.2872T,1998ApJ...503..325A}. 
In ATLAS we refer to this reference sky as the ATLAS wallpaper, 
and we currently rebuild this wallpaper about once per year. We store
the wallpaper in files of square degree size, and routines allow any
ATLAS pointing on the sky to trigger a reconstruction of a wallpaper 
reference image of the appropriate size and alignment. 
We use a  modified version of the image subtraction algorithm 
\texttt{HOTPANTS}
\citep{2015ascl.soft04004B}
to subtract the matched, convolved reference image from the target frame. 

The ATLAS reduction pipeline has a custom built point-spread-function fitting routine 
that runs on the difference images to produce flux measurements of all sources 
that are detected at 5$\sigma$ or more above the background noise. 
This routine (written in \texttt{C}) is called \texttt{tphot}, and is based
on the algorithms discussed in 
\cite{2011PASP..123...58T} 
and 
\cite{2013PASP..125..456S}. 
This produces one headed ASCII table per image (internally appended with 
\texttt{.ddc}), and the contents are summarised in Table\,\ref{tab:ddc}. 
Some of these quantities are used in an initial filtering process to remove the
bogus sources on the difference images (see Section\,\ref{sec:selection}). 

\begin{table}[]
    \centering
    \caption{The contents of a \texttt{.ddc} file for a difference image contains all detections (5$\sigma$ significance or greater), each with the following measured attributes}
    \begin{tabular}{ll}
    \hline\hline
    \texttt{.ddc} file entry  &  meaning \\\hline
    RA        & astrometrically calibrated RA in decimal degrees of centroid \\
    Dec       & astrometrically calibrated DEC in decimal degrees of centroid \\
    mag       & AB magnitude from PSF fit \\    
    dmag      & error in AB mag \\
    x         & x pixel on CCD of centroid  \\
    y         & y pixel on CCD of centroid  \\
    major     & PSF FWHM major axis [pixels] \\
    minor     & PSF FWHM minor axis [pixels] \\
    phi       &  angle of major axis, counter-clockwise from x axis [degrees] \\
    det       &  integer indicating fixed, free, trailed, streak, or negative flux analysis \\
    chi/N     &  reduced $\chi^2$ of the PSF fit  \\
    pvr       &  \texttt{vartest} probability of source being a variable star \\
    ptr       &  \texttt{vartest} probability of source being a moving object or astrophysical transient\\
    pno       &  \texttt{vartest} probability of source being just a noise fluctuation \\
    pbn       &  \texttt{vartest} probability of source being a vertically extended electronic ``burn'' artifact \\
    pcr       &  \texttt{vartest} probability of source being a cosmic ray hit \\
    pxt      &   \texttt{vartest} probability of source being an electronic crosstalk artifact\\
    psc      &   \texttt{vartest} probability of source being a star subtraction residual ``star scar'' \\
    pmv       &  \texttt{vartest} probability of source being a moving object (subset of \texttt{ptr}) \\
    pkn       &  \texttt{vartest} probability of source being a known minor planet \\
    dup      &   integer indicating whether this PSF measurement should be kept or culled\\
    WPflx    &   forced flux measured in the wallpaper (reference sky) at the position of the transient \\
    dflx     &   error in  WPflx   \\
    \hline
    \end{tabular}
    \label{tab:ddc}
\end{table}

\begin{figure}
    \centering
    \includegraphics[width=6.7cm]{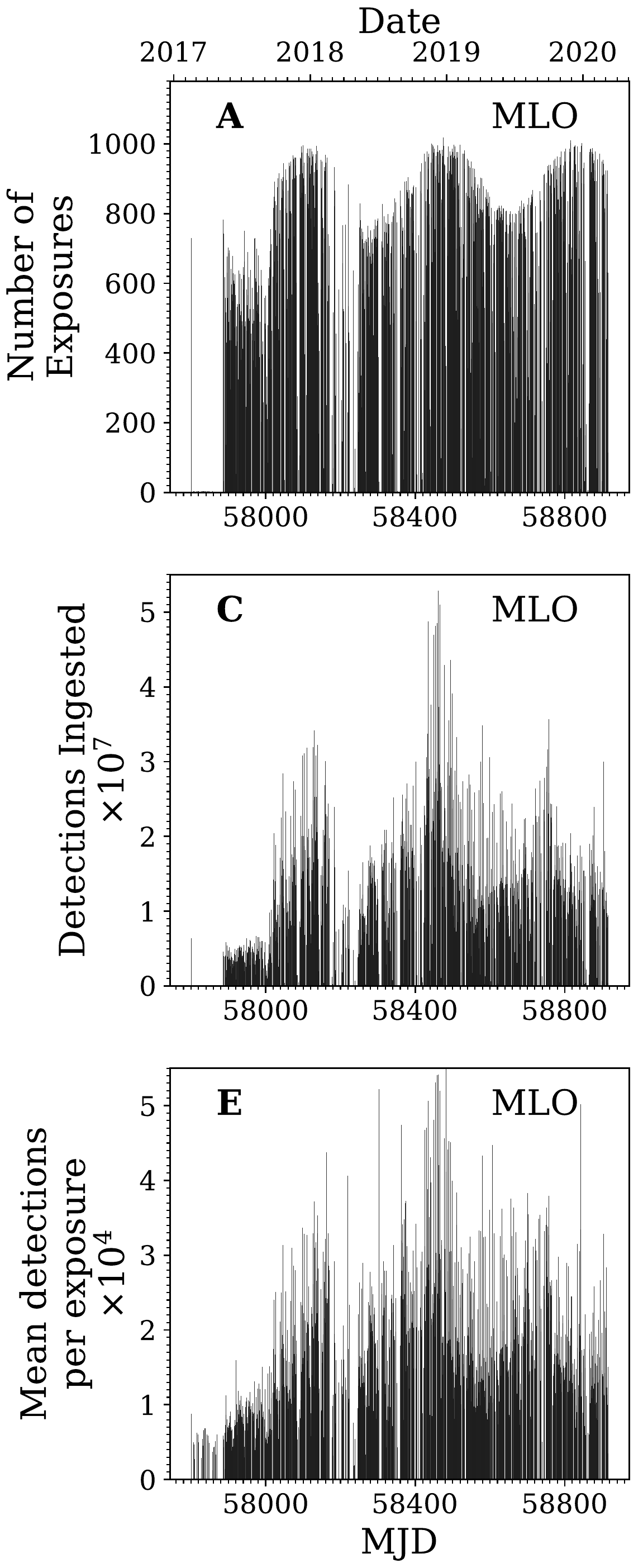}
    \includegraphics[width=8.547cm]{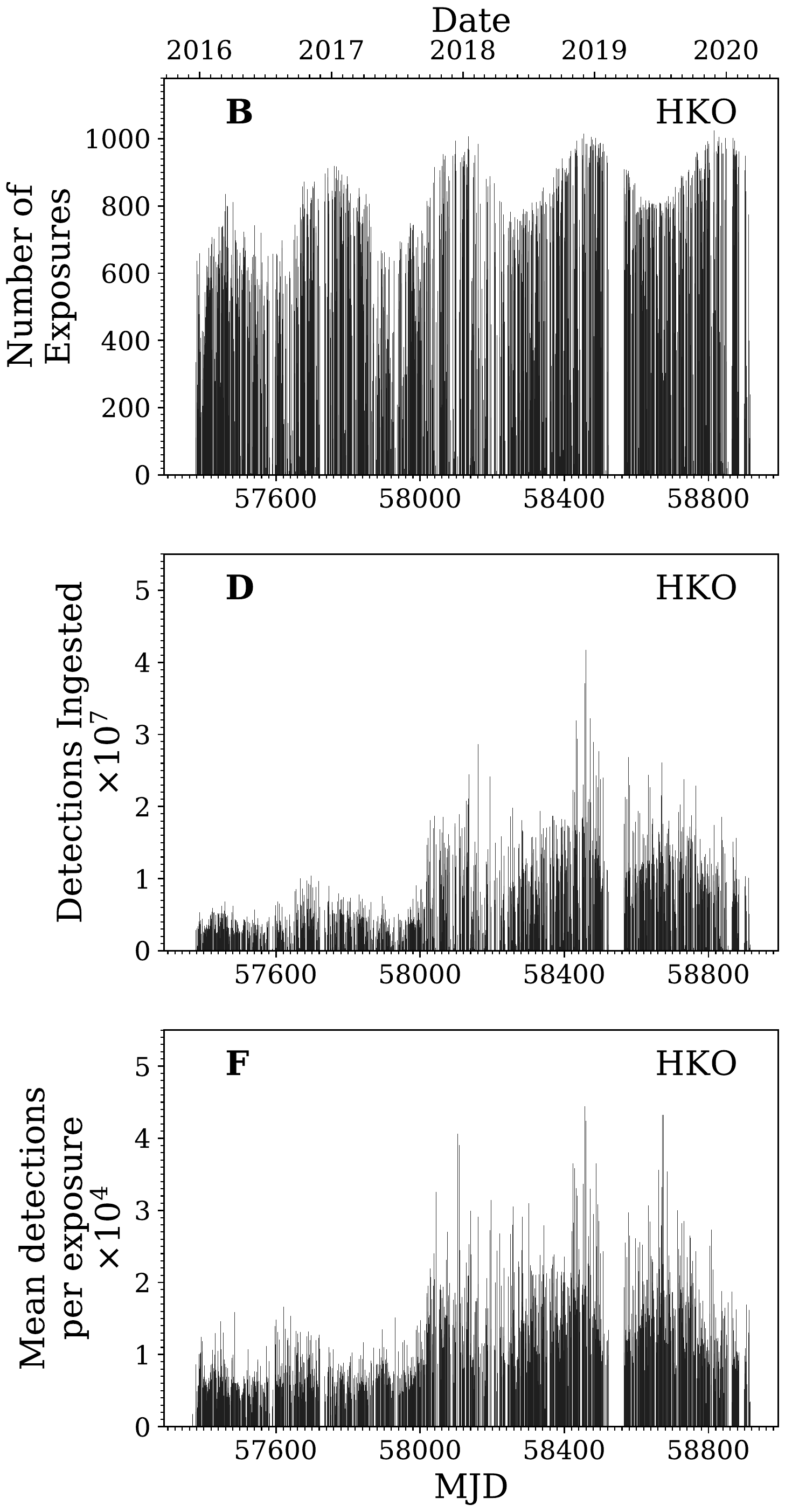}
\caption{{\bf A, C, E:} Number of exposures, number of detections ingested and average number of detections per exposure for MLO.  
The detections are the number of sources at 5$\sigma$ or greater that are present on the difference images (in the \texttt{.ddc} files). 
{\bf B, D, F:} The same for HKO.  The gap in the HKO data was caused by an ice storm on Haleakala.  See also Figure\,\ref{fig:monitor_new_kws}.  The rise in the average number of objects per exposure around MJD=58000 is because of a change of detection file schema and improvements in source extraction.   
}
\label{fig:Ingestion}
\end{figure}

On a typical night, a single ATLAS telescope produces about 
900 
catalogue files (one per exposure). 
Each catalogue file contains on average about 
20\,000 detections, 
so about 
10-20 million 
detections are ingested per night from the two telescopes combined. 
The vast majority of 
these sources are not astrophysically real objects and a 
series of quality control procedures and filters, together with 
machine learning algorithms are run to reject the false positives and 
leave real transient objects. Figure\,\ref{fig:Ingestion} illustrates
the number of detections ingested from these \texttt{.ddc} files per night
and example 
breakdowns of their  classifications are in 
Table\,\ref{tab:ingestnumbers}.

\begin{figure}
    \centering
    \includegraphics[width=8cm]{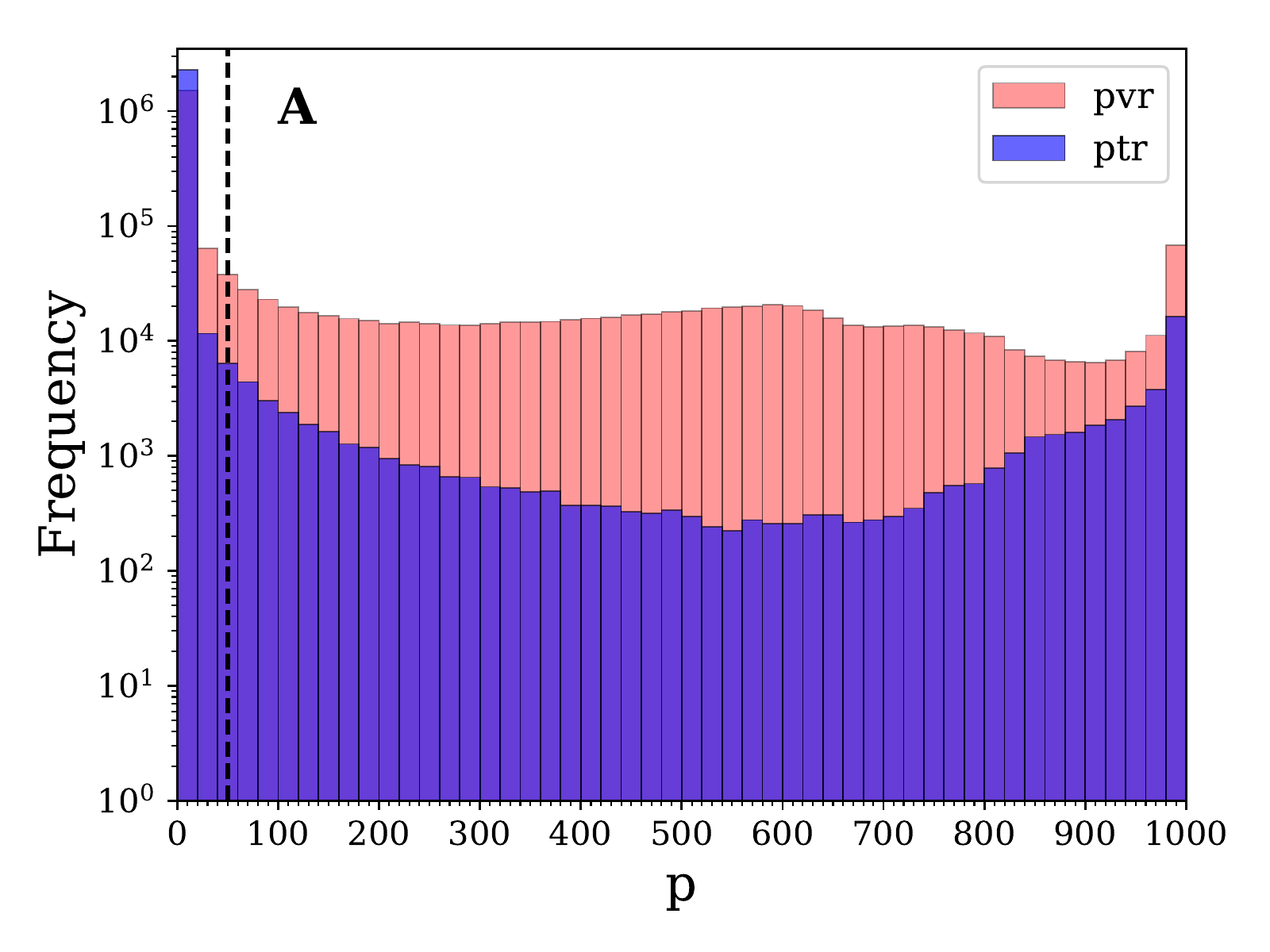}
    \includegraphics[width=8cm]{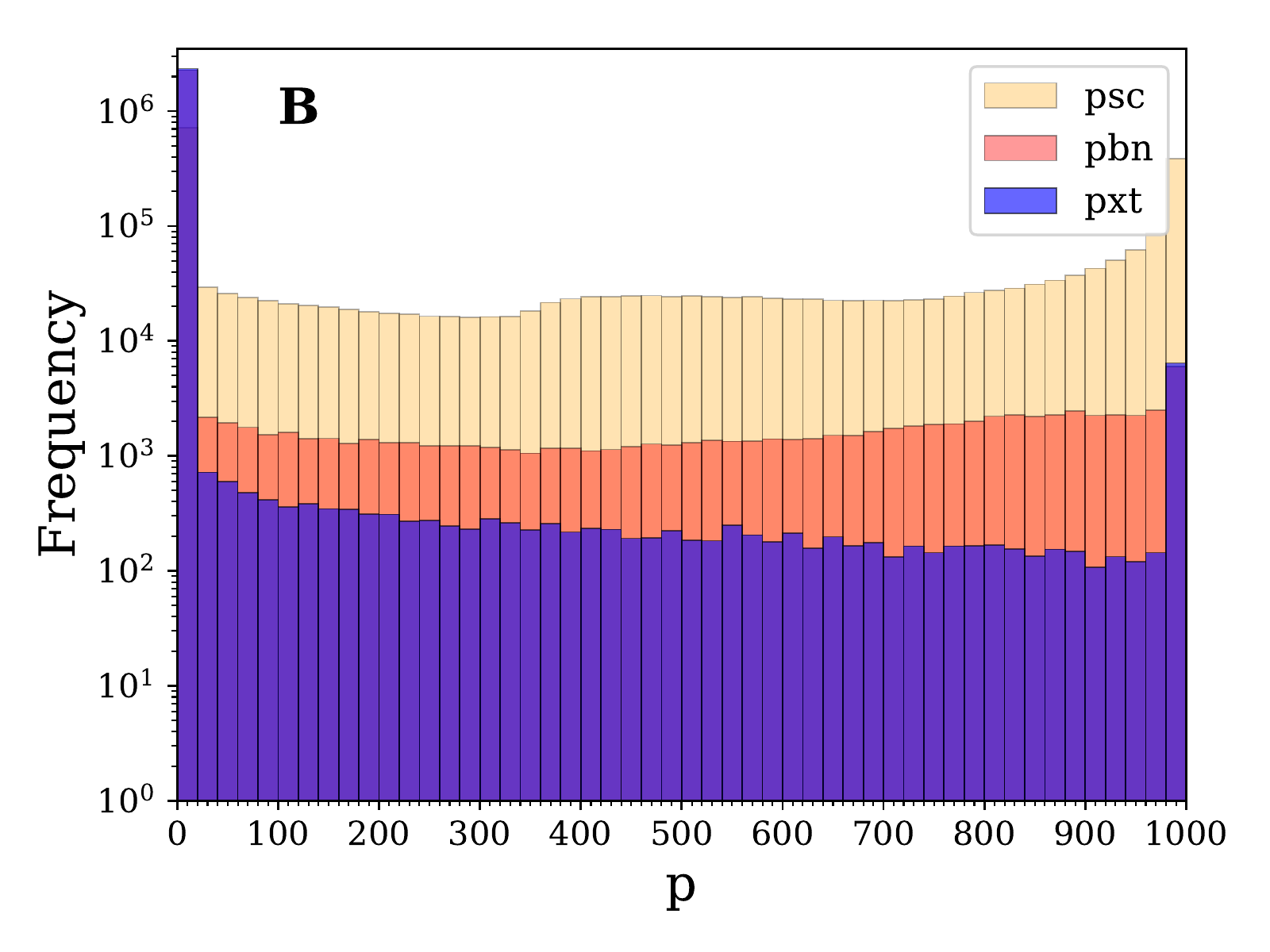}
    \includegraphics[width=8cm]{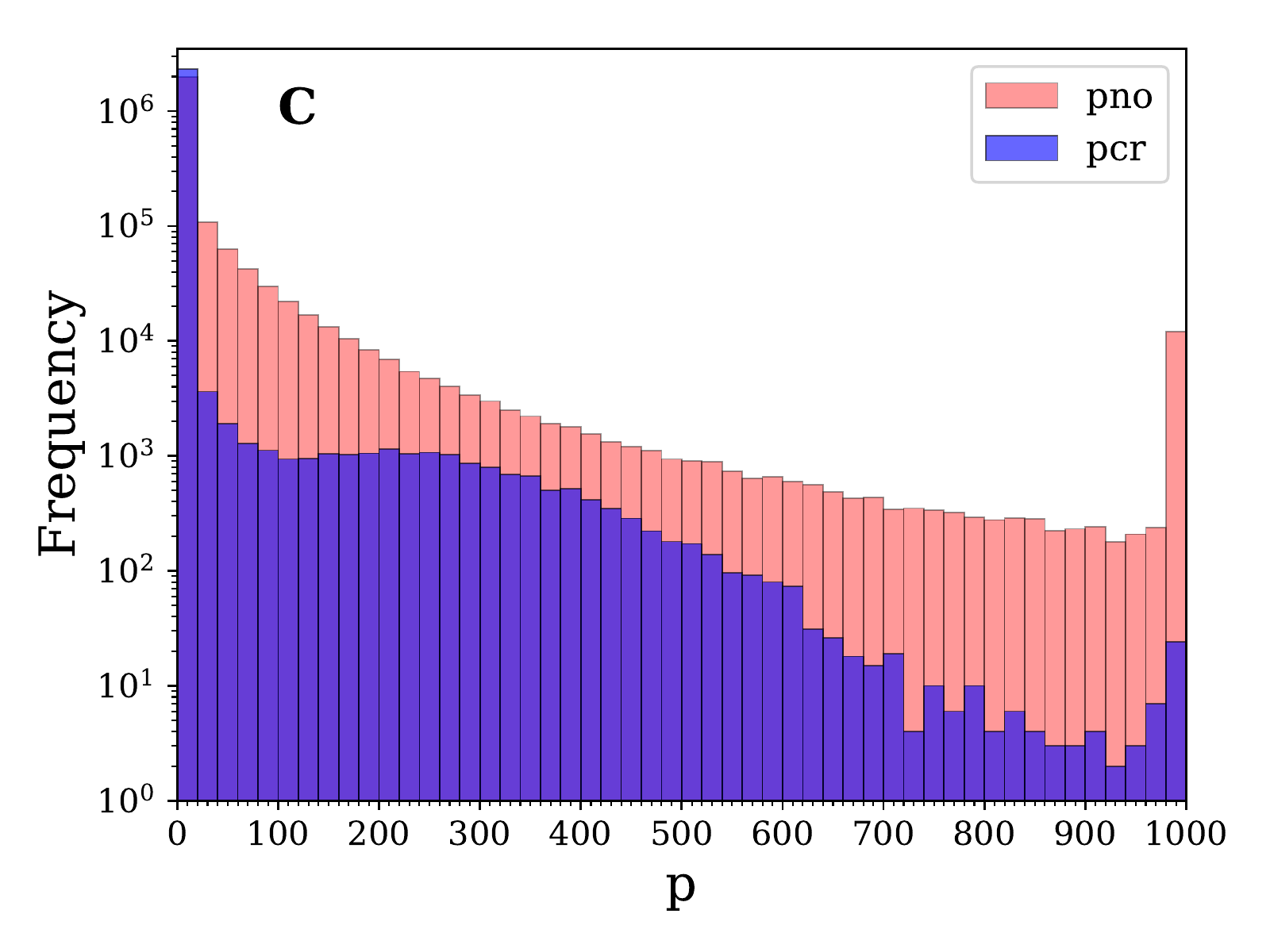}
    \includegraphics[width=8cm]{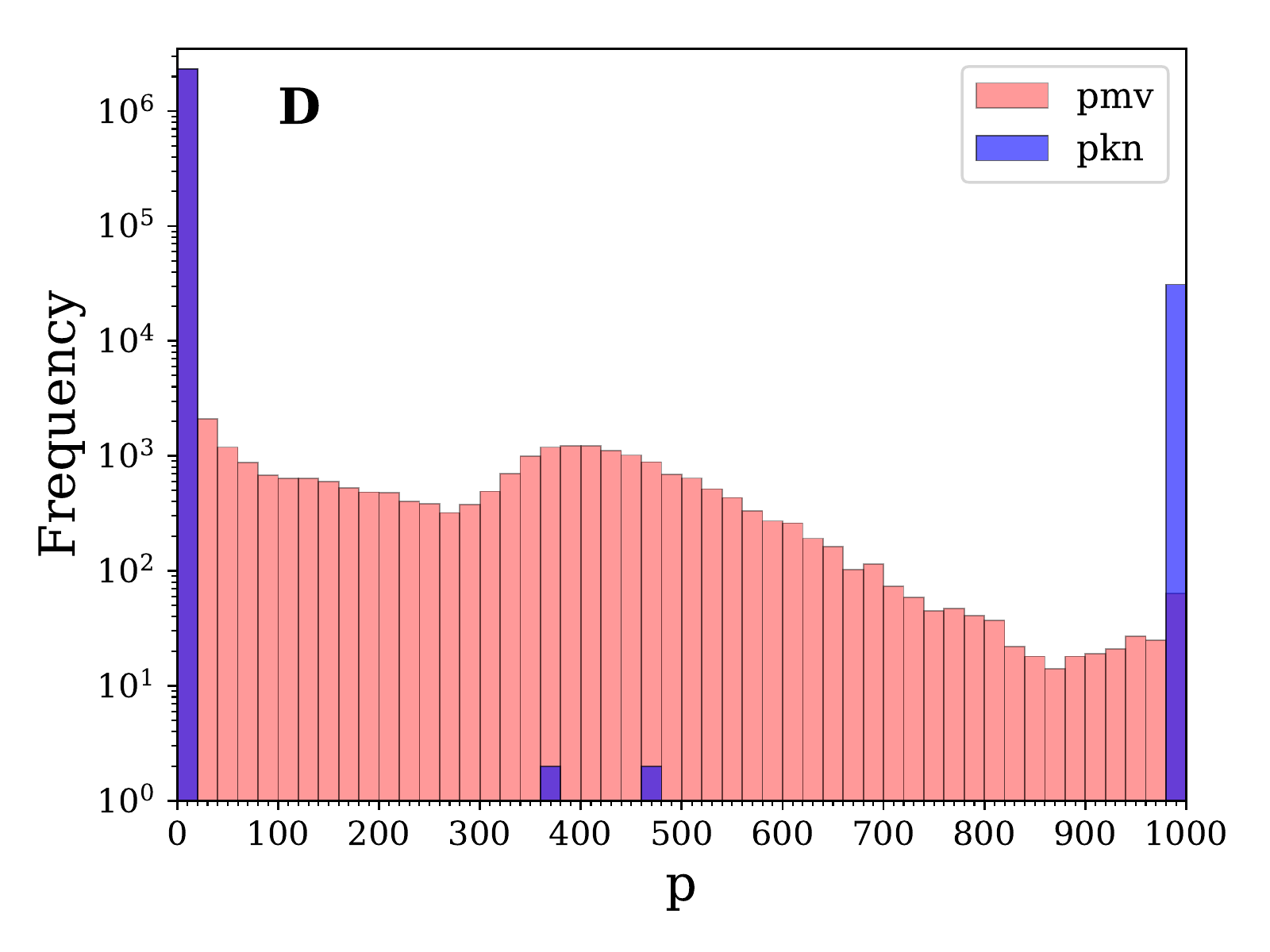}
\caption{Histograms of the \texttt{vartest} detection values for night of MJD=58816 (2019 November 29).  The output of \texttt{vartest} produces values for each parameter $0<\texttt{p}<999$. Of the 31 million detections that night, only those that form objects comprising at least three distinct intra night detections (2.4 million detections, 602\,000 objects) are shown.
{\bf A:} Histogram of the values of \texttt{pvr} and \texttt{ptr}.  The dotted line is the current threshold above which a detection is regarded as good quality for the purpose of detecting variable objects and transients. 
{\bf B:} Histogram of the values of \texttt{psc}, \texttt{pbn} and \texttt{pxt}.
{\bf C:} Histogram of the values of \texttt{pno} and \texttt{pcr}.
{\bf D:} The values of \texttt{pkn} and \texttt{pmv} are treated separately and the histogram of values is plotted.  The value of \texttt{pkn} is normally 999 or 0 (i.e. a detection is associated with a known object or it is not), whereas the value of \texttt{pmv} will span the entire range. Of 160\,000 known minor planet detections that night, 31\,000 detections were of movers slow enough to be aggregated into 4\,800 objects comprising 3 or more detections.
}
\label{fig:VartestHistograms}
\end{figure}

\section{The ATLAS Transient Science Server}
\label{sec:trans-server}

\subsection{Object definition and spatial indexing} 
\label{sec:objectdef}
The essential functions of a hardware and software system to deliver 
real transients from a synoptic wide-field survey such as ATLAS includes
amalgamating detections into objects, removing false positives, 
assimilating lightcurves, and associating the newly discovered objects
with known catalogues. We have built the ATLAS Transient Science Server
to provide all these functions.  The Server is currently based on individual detections from single ATLAS exposures, but in future we intend to combine the information from the nightly difference stacks to allow us to detect transient objects earlier.

The ASCII catalogue files of objects detected on the difference 
images (the \texttt{.ddc} files described in Section\,\ref{sec:telsurvey}) are transferred to a machine at Queen's University Belfast (QUB). This is a 28 core server with 128GB RAM (also used as the ingestion machine).
The files are transferred (using parallel download threads) in the background every 30 minutes between 04:00 UT and 22:00 UT (corresponding to the earliest start of a Hawaiian observing night at 18:00 and cleaning up after the night is done at 12:00 next day). 
Just before an ingest cycle begins, an extra check is done to make sure we have the very latest dataset. The reduced, calibrated images and their associated subtracted frames are also transferred (also in parallel) to QUB during the Hawaiian night.  Typical end to end data rates from the University of Hawaii Manoa campus to the QUB machine are 40 MBytes/sec.  Since individual download threads are throttled by the network infrastructure, using 8 parallel threads ensures we use the maximum available bandwidth. Each of the 8 threads achieves 5 MBytes/sec and
using more than 8 threads brings very little gain in download speed.

The \texttt{.ddc} file contains a header comprising 37 key-value pairs describing metadata of the exposure, such as the telescope pointing coordinates, exposure MJD, exposure time, filter, PSF FWHM, zero point and $5\sigma$ limiting magnitude.  This is followed by a headed table with each entry corresponding to a single detection (see Table\,\ref{tab:ddc}).  The metadata and detections are automatically read by a \texttt{C++} ingester (generated by python code which reads the \texttt{.ddc} structure) at QUB which connects to a local MySQL database.  Detections are first checked via cone searching (see below) against previously ingested objects.  If a detection is within 3\farcs6 (twice the ATLAS pixel size) of an existing object, it is ingested and formally associated with that object, becoming another point on the lightcurve.  If there are no objects within the specified proximity of the ingested detection, a new object is created and the detection is associated with the new object.  To speed up the ingest, multiple ingester processes are run in parallel using a python wrapper (to utilise all the cores of the ingestion machine).

Hence all records in the \texttt{.ddc} files are preserved in the database. The critical values ingested are RA, Dec, calculated HTM ID (see below), mag and dmag.  Other detection properties are also ingested to facilitate post ingest cutting.

The detection data are spatially indexed using Hierarchical Triangular Mesh \citep{szalay2005indexing}. 
Since the vast majority of queries are cone searches of the order of a few arcsec in size, HTM Level 16 ($2\times4^{(16+1)}$ triangles, hence mean triangle area = 
15.6 square arcsec)
is used. The HTM value, calculated for every detection's RA and Dec upon ingest, is an integer and this column is indexed inside the relational database.  The \texttt{C++} HTM library provides a cone searching API which returns the ranges of triangles that overlap a cone on the sky.  This is used to create an SQL \texttt{WHERE} clause.  All objects in the database that have those triangle IDs are returned to the requesting code.  The cone search code then just needs to eliminate objects outside the requested radius (i.e. the edges of the resultant jagged cone are smoothed).

\subsection{Selection criteria, machine learning and classification} 
\label{sec:selection}

After ingest, selection criteria and algorithms are applied to select real astrophysical transients from the detection stream.

{\em Multiple detections:} since the standard quad set of 30\,sec images are separated across a 1\,hr window of observation, spatial coincidence of the sources is used to reject some moving objects and other non-astrophysical contaminants. Three or more good quality, co-spatial  
detections are required to define an object and all three must be on the same night.  They must also not be within 100 pixels of the edge of the detector. There is no selection applied to the coherence of the magnitudes of three detections, and we do 
find rapidly evolving transients (over the course of 1 hour), all of which we have identified as M-dwarf flares apart from one 
GRB afterglow \citep{2017ApJ...850..149S}.

The main benefit of reducing the triggering threshold to 2 detections rather than 3 would be that we may discover and detect objects earlier. Given the 2 day cadence of ATLAS, objects with only 2 detections on a single night and no other 5$\sigma$ detections are either real transients that just rise to the sensitivity limit in that window, earlier detections of objects that continue to rise (and are subsequently detected with $\geq3$ detections afterwards) or spurious artefacts.  We carefully checked 50 real objects from 2020 April, and found that 40\% could potentially have been discovered earlier had we used a 2 detection minimum criterion rather than 3. The mean time difference was a gain of 4 days. However while that means 40\% of objects have the potential to have been discovered earlier, it does not mean they would definitely have been promoted. The early detections would need both a high RB score (see below) and to be confidently judged by human scanners as real. Our human scanners will often wait for an extra night of data if the images are ambiguous. This gives a sense of the maximum gain if we were to go to a criterion of 2 images. It is possible, but the penalty is a large increase in false positives.  For example, the figures shown in brackets in Table\,\ref{tab:ingestnumbers} illustrate that human scanners would have been presented with over 24 times more objects. Seven objects submitted to the IAU TNS by other surveys (PS1 and ZTF), all fainter than 19 mag at peak, would have been triggered with a 2 detection threshold by ATLAS, all with a high RB factor, but {\em after} the other surveys had already reported them.

{\em Source classification: } A pixel-based classification of candidate source detections in the ATLAS difference images is performed using a program called \texttt{vartest}, written in \texttt{C} (Heinze et al. in prep), which runs in Hawaii after the difference image detections have been extracted.  \texttt{vartest} attempts to classify sources 
into several categories of real and spurious objects. It is not a machine learning code, but rather uses a logical cascade of manually tuned thresholds and analytical probability estimators applied to various pixel-based metrics. These metrics include centroid offset and flux ratio between the original and difference images; elongation of the PSF; pixel-coordinate association with bright stars; ratio of nearby bright and dark pixels; and many others. The classification is done probabilistically, with $p=0$ being the lowest probability and $p=999$ being the highest. The categories of real objects are known asteroids (\texttt{pkn}), unknown asteroids or transients (\texttt{ptr}), and variable stars (\texttt{pvr}). Objects with nonzero values of \texttt{ptr} are also assigned a probability of being a fast-moving (i.e. trailed) detection (\texttt{pmv}). The categories of spurious detections are noise (\texttt{pno}), electronic column artifact or ``burn'' (\texttt{pbn}), cosmic ray (\texttt{pcr}), electronic crosstalk (\texttt{pxt}), and star subtraction residual or ``star scar'' (\texttt{psc}).
These are summarised in Table\,\ref{tab:ddc}. The sum of the values for a single detection (excluding \texttt{pmv} and \texttt{pkn}, which are subsets of \texttt{ptr}) will not exceed 999.  \texttt{vartest} also identifies duplicate detections in its input file, and attempts to indicate the ``best'' detection (stored in the \texttt{.ddc} \texttt{dup} column).

From the \texttt{vartest} set of parameters we define potential ``good quality'' detections as those with values $\texttt{pvr} > 50$ or $\texttt{ptr} > 50$. If either of these two criteria are not met, the detection does not count toward the  3 minimum good quality detections within a night required to define an object.  Additionally, the detection must NOT belong to a known mover ($\texttt{pkn} < 500$) or a possible mover ($\texttt{pmv} < 500$) or be marked as crosstalk ($\texttt{pxt} < 500$).  The value of \texttt{det} must not indicate negative flux and the value of \texttt{dup} must not indicate a deprecated duplicate value.  Figure\,\ref{fig:VartestHistograms} shows histograms of the vector values for the full (dark) night of MJD=58816 (2019 November 29).  The data rows used in the plots are only those which comprised objects of 3 or more distinct detections.  The dotted line on the \texttt{ptr} and \texttt{pvr} plot shows where the quality cut is currently set.

{\em Object recentering:}
If an object passes the above criteria, the detection closest to the mean coordinates of the object is identified as the representative detection for that object.  This ensures that subsequent detections are associated with the correct objects. (The code is periodically rerun to make sure the object is properly centred.)

{\em Ephemeris checker:}
An initial cross-match with known minor planets is done in Hawaii (within \texttt{vartest}) 
using a local copy of the Minor Planet Center (MPC) orbital elements and a 9.3\,arcsecond (5 pixel) association 
radius. This produces an entry in the \texttt{.ddc}, labelled as \texttt{pkn} which is the probability that the detection is that of a known asteroid.  
 However, this occasionally misses some asteroids, and does not currently label any comets and thus we implement a further check.  Three times weekly, at QUB, the Minor Planet Center (MPC) orbital element database is downloaded and converted into an XEphem database file.  At the same time, the comet elements, provided by the MPC in XEphem format already, are appended to the converted downloaded file.  After each batch of data is ingested, the latest detection of an object that passes previous cuts is passed to code which locally generates the solar system object ephemeris from the orbital element data based on the MJD of the exposure.  The detections are grouped by MJD to minimise the number of times an ephemeris needs to be generated.  A cone search is then run against the ephemeris to identify any solar system object (asteroid or comet) that may lie within a specified search radius (currently 10 arcsec).  If it lies within the search radius, the object is tagged as a probable known mover and labelled with the solar system object ID.  The ephemeris check is done offline without interrogating the MPC servers (apart from the periodic data downloads) to minimise MPC load.
Additionally, moons of the giant planets are not currently recorded in the MPC database of orbital elements and therefore fail to be found in this basic check.  For this reason, the positions of all the satellites of the giant planets have been generated\footnote{via Giorgini, JD and JPL Solar System Dynamics Group, NASA/JPL Horizons On-Line Ephemeris System, \url{http://ssd.jpl.nasa.gov/?horizons}, data retrieved 2019-02-21.} for every hour for the next 10 years and these are stored in a local database table at QUB.  A simple cone search is run against this for the most recent detection of each object, with an additional temporal check of any matched object.  If a matched planetary satellite is within 20 arcsec and half a day of the timestamp of the detection, the object is labelled as a probable mover.

Although every effort is made to reject moving objects from the flow of data to human scanners, two serendipitous solar system discoveries have been made with the ATLAS Transient Science Server. We discovered unusual activity of the asteroid (6478) Gault \citep{2019CBET.4594....1S,Kleyna_2019}, arising from a comet-like ``tail''  which provided a locus of detections offset from the asteroid itself. We also discovered a distant  previously unknown long period comet C/2019 K7 (Smith) \citep{2019CBET.4645....1S,2019CBET.4648....1S}, at 4.5AU perihelion which was therefore
slow moving and made its way through the Transient Science Server as a possible stationary transient. 

{\em Sherlock:} Once an object is defined, a boosted decision tree algorithm
(internally known as \emph{Sherlock}) mines a library of historical and on-going
astronomical survey data and attempts to predict the nature of the
object based on the resulting crossmatched associations found. 
One of 
the main purposes of this is to identify variable stars, since they make up about 
50\% of the objects  (see Table\,\ref{tab:ingestnumbers}), and to associate candidate extragalactic 
sources with potential host galaxies.  The full details of 
this general purpose algorithm and its implementation will be presented in an upcoming paper 
(Young et al. in prep), and we give an outline of the algorithm as implemented
within the ATLAS Transient Science Server here. 

The library of catalogues contains datasets from many 
all-sky surveys such as the major Gaia DR1 and DR2 \citep{2016A&A...595A...2G,2018A&A...616A...1G},
the Pan-STARRS1 Science Consortium surveys 
\citep{2016arXiv161205560C,magnier2017a,magnier2017b,magnier2017c,flewelling2017} and 
the catalogue  of probabilistic classifications of unresolved point sources by 
\cite{2018PASP..130l8001T} which is based on the Pan-STARRS1 survey data. 
The  SDSS DR12 PhotoObjAll Table, SDSS
DR12 SpecObjAll Table \citep{2015ApJS..219...12A} usefully contains both 
reliable star-galaxy separation and photometric redshifts which are 
useful in transient source classification. 
Extensive catalogues with lesser spatial resolution or colour information that 
we use are  the GSC v2.3 \citep{2008AJ....136..735L} and 2MASS catalogues \citep{2006AJ....131.1163S}. \emph{Sherlock} employs many smaller
source-specific catalogues such as Million Quasars Catalog v5.2 \citep{2019arXiv191205614F}, Veron-Cett AGN
Catalogue v13 \citep{2010A&A...518A..10V}, Downes Catalog of CVs \citep{2001PASP..113..764D}, Ritter Cataclysmic Binaries
Catalog v7.21 \citep{2003A&A...404..301R}. For spectroscopic redshifts 
we use the GLADE Galaxy Catalogue v2.3 \citep{2018MNRAS.479.2374D} and
the NED-D Galaxy Catalogue
v13.1\footnote{https://ned.ipac.caltech.edu/Library/Distances/}. \emph{Sherlock} 
also has the ability to remotely query the NASA/IPAC
Extragalactic Database, caching results locally to speed up future
searches targeting the same region of sky, and in this way we have built up an
almost complete local copy of the NED catalogue. 
More catalogues are
continually being added to the library as they are published and become
publicly available.

 At a base-level of matching Sherlock distinguishes between transient objects \emph{synonymous} with (the same as, or very closely linked, to) and those it deems as merely \emph{associated} with the catalogued source. The resulting classifications are tagged as \emph{synonyms} and \emph{associations}, with synonyms providing intrinsically more secure transient nature predictions than associations. For example, an object arising from a variable star flux 
 variation would be labeled as \emph{synonymous} with its host star since it would be astrometrically coincident
 (assuming no proper motion) with the catalogued source. Whereas an extragalactic supernova would typically be 
 \emph{associated} with its host galaxy - offset from the core, but close enough to be physically associated.
 Depending on the underpinning characteristics of the source, there are 7 types of predicted-nature classifications
that Sherlock will assign to a transient:

\begin{enumerate}

\item
  \textbf{Variable Star} (VS) if the transient lies within the synonym
  radius of a catalogued point-source,
\item
  \textbf{Cataclysmic Variable} (CV) if the transient lies within the
  synonym radius of a catalogued CV,
\item
  \textbf{Bright Star} (BS) if the transient is not matched against the
  synonym radius of a star but is associated within the
  magnitude-dependent association radius,
\item
  \textbf{Active Galactic Nucleus} (AGN) if the transient falls within the
  synonym radius of catalogued AGN or QSO. 
\item
  \textbf{Nuclear Transient} (NT) if the transient falls within the
  synonym radius of the core of a resolved galaxy,
\item
  \textbf{Supernova} (SN) if the transient is not classified as an NT
  but is found within the magnitude-, morphology- or distance-dependant
  association radius of a galaxy, or
\item
  \textbf{Orphan} if the transient fails to be matched against any
  catalogued source.
\end{enumerate}

For ATLAS the synonym radius is set at $1.5''$. This is the
crossmatch-radius used to assign predictions of VS, CV, AGN and NT.
The process of attempting to associate a transient with a catalogued
galaxy is relatively nuanced compared with other crossmatches as there
are often a variety of data assigned to the galaxy that help to greater
inform the decision to associate the transient with the galaxy or not.
The location of the core of each galaxy is recorded so we will always be
able to calculate the angular separation between the transient and the
galaxy. However we may also have measurements of the galaxy morphology
including the angular size of its semi-major axis. For ATLAS we reject
associations if a transient is separated more than 2.4 times the
semi-major axis from the galaxy, if the semi-major axis measurement is 
available for a galaxy. 
We may also have a distance measurement or redshift 
for the galaxy enabling us to convert angular separations between
transients and galaxies to (projected) physical-distance separations. If a transient
is found more than 50 Kpc from a galaxy core the association is
rejected.

Once each transient has a set of independently crossmatched synonyms and
associations, we need to self-crossmatch these and select the most
likely classification. The details of this will be presented in a 
future paper (Young et al. in prep). Finally the last step is to 
calculate some value added parameters for the transients, such as
absolute peak magnitude if a distance can be assigned from a matched
catalogued source, and the predicted nature of each 
transient is presented to the user along with the 
lightcurve and other information (see Figure\,\ref{fig:webcandidatepage2019tua}).

We have constructed a 
 multi-billion row  database which contains all these catalogues. It 
 currently consumes about 4.5TB and sits on a separate, similarly specified machine to that of the ATLAS database.  It will grow significantly as new catalogues are added (e.g. Pan-STARRS 3$\pi$ DR2, VST and VISTA surveys, future Gaia releases etc).

{\em Cross-matching with other surveys:}  The flagged objects are also cross-matched against a database of transient objects discovered by other surveys and tagged if they are already known. 
We compile a local database from the IAU TNS and
also cross-match against objects from ZTF \citep{2019PASP..131a8002B}, which are
contained in the Lasair broker and are likely to be supernova or extragalactic transients of some sort \citep{2019RNAAS...3...26S}. 
Hence users know almost instantaneously if an object has been discovered by another survey and registered in the TNS as a discovery, or 
if it has multiple detections in the ZTF public stream.  This crossmatch runs at the end of every ingestion cycle (at least 4 times a day) with a match radius of 3\farcs0.

We also trawl ``The Astronomer's Telegram''\footnote{http://www.astronomerstelegram.org} text reports (ATels) for information connected to ATLAS transients. There are two distinct searches performed against the human-written, plain-text content found in the title and body of each ATel; a name-search and a coordinate-search. As ATLAS and all other major transient search projects have consistent transient naming schemes, we have found it possible to compile a set of regular-expressions that match most of the transient names reported in the ATels. All matched names are recorded in a separate table to the original database table hosting the ATel titles and content, with ATel number as the relational key between tables. Furthermore a long and complex regular expression is employed to extract out all sets of equatorial coordinates found within ATels. This expression has to accommodate for the plethora of formats of sexagesimal coordinates and also for the surprisingly common human-errors found in those reports. The coordinates are homogenised into decimal degree format and recorded in another database table which is then indexed using the familiar HTM indexing scheme. Whenever a new ATel is released, transient names and coordinates can then be matched against ATLAS transient names, pseudonyms from other surveys and coordinates. The discovery of a new ATLAS transient also triggers a name and coordinate-search for matches in past ATels. The resulting ATel matches are reported on the ATLAS transient pages providing direct web-links to the associated ATel reports.
However use of ``The Astronomer's Telegram'' for registering, announcing and classifying extragalactic transients and supernovae
is dwindling, being replaced by a more robust and accessible database system in the IAU TNS. The latter allows full photometric 
and spectroscopic registration in an automated way, combined with a well maintained database and links to WISeREP \citep{2012PASP..124..668Y}. It also allows reports to be written and released, as AstroNotes \citep{2019TNSAN...1....1G}, all of which are indexed on the NASA Astrophysics Data System and are citable, hence providing all the functionality of ATels but with added data curation and searching value.

\begin{figure}
    \centering
    \includegraphics[width=8cm]{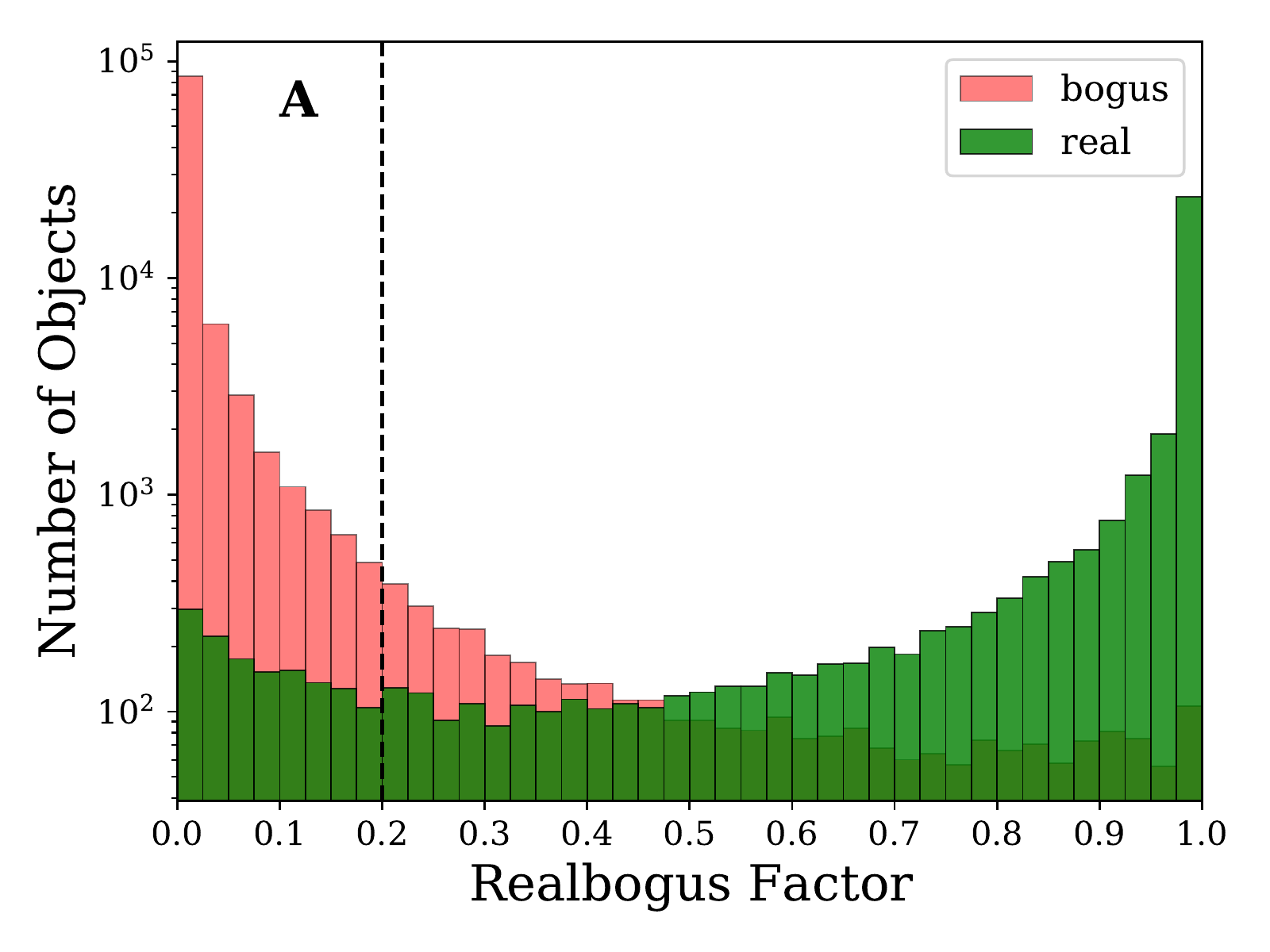}
    \includegraphics[width=8cm]{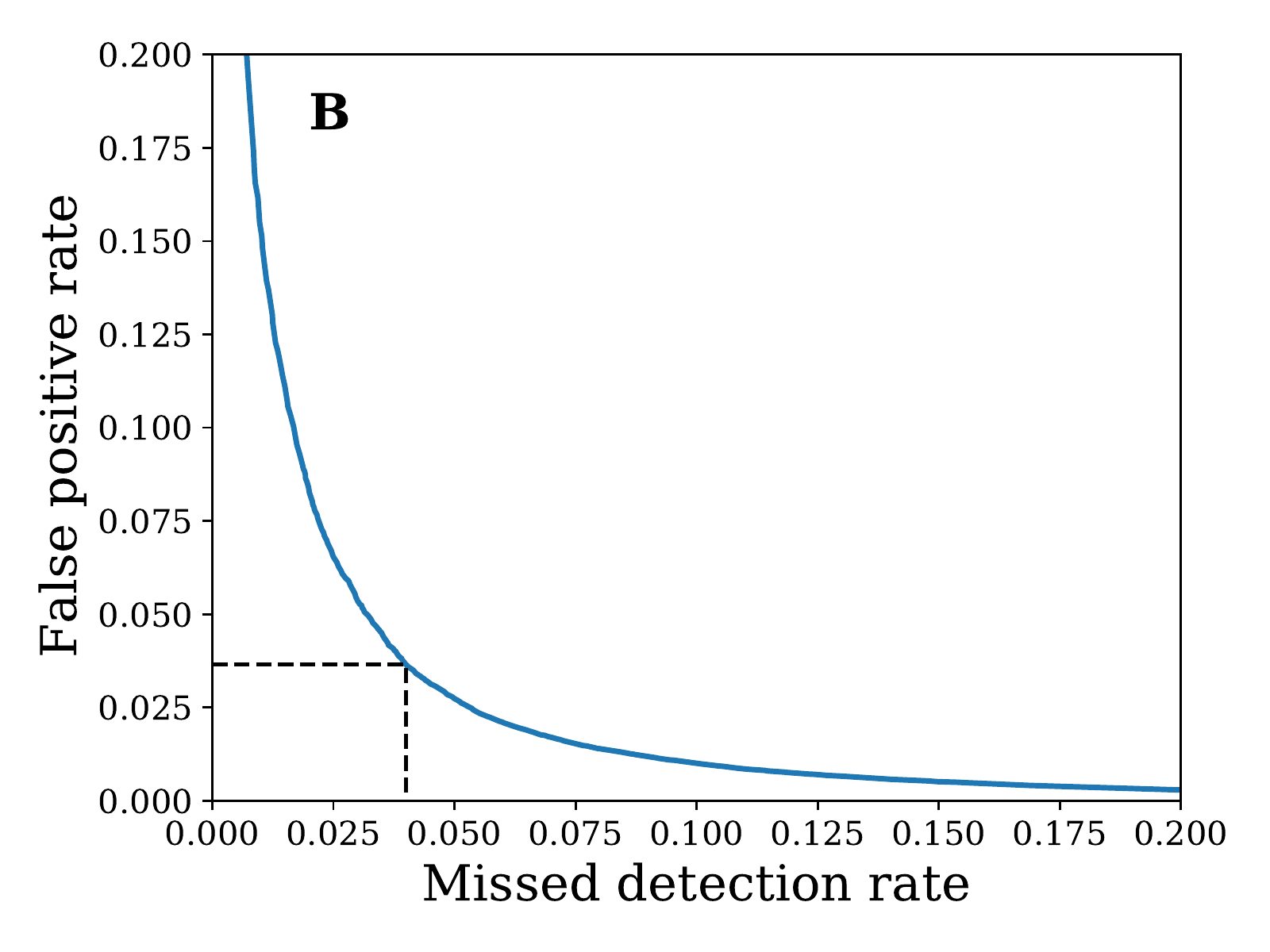}
\caption{
{\bf A:} Histogram showing the classifier performance against the training set (for the Haleakala (HKO) classifier). The dotted line shows the RB factor threshold which corresponds to a 4\% missed detection rate of real 
objects.  The real/bogus value threshold is $RB = 0.2$, below which objects are regarded as bogus. (Note the logarithmic vertical scale.)
{\bf B: } Detection vs Error tradeoff. The dotted line shows the false positive rate (3.65\%) at the set missed detection rate of 4\%, based on a test subset of the training set. This is for an $RB = 0.2$ threshold for real/bogus selection.   }
\label{fig:ROC}
\end{figure}

{\em Machine learning image recognition:} For the objects that 
remain after the stars are removed, we construct a 
triplet of ``postage stamps''. The target image, reference and difference images centred on the detections are produced of size 6.2 arcmin (200$\times$200 pixels). 
A machine learning algorithm, trained to recognise ``good'' subtractions (i.e. PSF-like objects in the centre of the difference stamp) is run on the multiple (at least three) difference images associated with each object.  A median score between 0 (bogus) and 1 (real) is applied to each object.  All objects with a score above a  chosen threshold are passed to human scanners.  The training sets were built from the 20$\times$20 pixel core of difference image stamps centred on 200\,000 known asteroids (real) from data sampled over six separate nights and a random selection of the 20$\times$20 cores of difference image stamps of 600\,000 objects that were not known variables (bogus).  Two training sets were built, one for each telescope, both of which have similar performance, though the HKO dataset was built using three cyan and three orange nights, whereas the MLO classifier only uses orange nights (because it only has the one filter).  Figure\,\ref{fig:ROC} shows the performance of the classifier, displaying a plot similar to a receiver operator characteristic (ROC) curve and a hypothesis distribution.  We began using the
 machine learning algorithm as 
described in \cite{Wright2017} and \cite{Wright15} (random forest and sparse filtered neural network)
but now use a Keras \citep{chollet2015keras} 
convolutional neural network with a Tensorflow \citep{tensorflow2015-whitepaper} backend. 
Convolutional neural networks (CNNs) were  applied successfully for transient discovery and detection in image data by \cite{2017ApJ...836...97C}, showing excellent promise for use in high data volume surveys
\citep[see also][]{2018arXiv180803626R}. Our Keras/Tensorflow 
CNN, which is a modification of an example model used to recognise the MNIST database of handwriting digits, consists of three 2D convolutional layers with 16, 32 and 64 filters and a kernel size of 2$\times$2 pixels.  Each convolutional layer is followed by a maxpooling layer that pools over 2$\times$2 pixel regions. The final pooling layer feeds into a dropout layer (to prevent overfitting) with a dropout rate of 0.3, which is followed by a Dense (fully connected) layer with 500 units and another dropout layer with a dropout rate of 0.4. The output layer is a Softmax classifier that produces two outputs; the probability the detection is real and the probability the detection is bogus. All units other than those in the output layer are rectified linear units (relu). ``Same'' padding \citep{dumoulin2016guide_same} is used throughout and the model is trained with the categorical cross entropy loss \citep{hastie2009elements_cross_entropy} and optimised with adam \citep{kingma2014adam}.

We are of course free to set the threshold for the RB factor to any value to optimise either completeness or purity. In reality we run our 
daily transient searches by selecting objects with $RB \geq 0.2$, in the knowledge that this results in 96\% completeness. As can be seen in the example numbers in Table\,\ref{tab:ingestnumbers}, this reduces the objects that humans will scan by a factor 20, from about 9000 to 300, leaving quite a manageable number per day. The classifier is periodically retrained as the training sets are refined.  This is necessary because of changing conditions (e.g. PSF improvements over time).  Selection of representative ``bogus'' data is also improving as human scanners log more examples where machine classification has failed. 
We foresee improvements in the training sample by using all the images of all objects
we now consider as real stationary transients. This may improve the performance of the 
classifier on highly varying spatial backgrounds (which leave noise residuals in the
difference images). With close to 8000 real objects in our database and at least 
20 images per object this will provide a rich and useful training set. 
Employing an unsupervised method could allow clusters of real and bogus 
objects to emerge and further improve performance.

{\em Forced photometry:}  For each potential extragalactic transient, or orphan transient, 
that has minimum of 3 detections  at 5$\sigma$ significance, and a machine learning score above our chosen threshold (i.e. has passed the above filters), we carry out forced photometry
on historic images up to 30 days before initial discovery at the position of the transient (using the
rolling average of its RA and DEC in the unforced \texttt{.ddc} files).
This normally results in one or more pre-discovery epochs
with multiple 2-4$\sigma$ detections on the individual 30 sec frames.
We calibrate and measure the forced photometry in flux ($\mu$Jansky),
since conversion between AB magnitudes is trivial, allowing low
significance positive values, zero values, and negative values to provide
real meaning (which is lost when converting to Pogson magnitudes). 
This is invaluable in constraining the explosion epoch for nearby
objects and providing constraining early data for all transients. This
information is presented to the user along with all the above information, as 
explained in the next section (see Figure\,\ref{fig:webcandidatepage2019tua} for an example). 
The forced photometry is again PSF based, 
with \texttt{tphot} used in the same way as for unforced PSF flux measurements, 
but simply with the centroid position fixed and the PSF modeled 
from the input image before subtraction. Forced photometry is also done against a stack of the four intra-night difference images for an object, which can help to tie down the discovery epoch for an object.

\begin{table*}
\centering
\caption{Example nightly ingest numbers for two ATLAS telescopes on the night of MJD=58816 (2019 November 29).  Good quality detections are defined as having $\texttt{pvr} > 50$ or $\texttt{ptr} > 50$ and not a known or possible mover, must be positive flux and not crosstalk.  The detection must also not be within 100 pixels of the edge of the detector.  Note that the number of ``good'' objects for this night is higher than average. This was possibly because of dark time, the slightly longer winter night, and a full night of good quality difference images. The figures assuming 2 spatially coincident detections rather than 3 are also shown in brackets.
}
\label{tab:ingestnumbers}
\begin{tabular}{p{0.5\linewidth}rr}\hline 
Stage & Number &  (2 detection Number) \\
\hline
Detections on the difference images ($\geq5\sigma$ in the \texttt{.ddc} files) & $3.1\times10^{7}$  &   \\
Single detection objects (probably noise, edge, streaks and $1.6\times10^{5}$ known movers) &  $2.1\times10^{7}$  &   \\
Objects with 3 (2) or more spatially coincident detections  &  $6.2\times10^{5}$  &   {\bf ($2.6\times10^{6}$)} \\
Objects with 3 (2) or more spatially coincident, good quality detections  &  $2.2\times10^{4}$  &  {\bf ($3.6\times10^{5}$)} \\
Objects remaining after stars removed  &    9\,450  &  (68\,542) \\
Objects remaining after AGN removed  &    9\,300  &  (68\,068) \\
Objects remaining after untagged known slow movers removed (e.g. comets)  &  9\,280 & (68\,008) \\
Objects remaining after ML and $RB\geq0.2$ applied  &  340  &  (8\,341) \\
Good objects after human scanning &   40  &   \\
\hline 
\end{tabular}
\end{table*}

\subsection{User interaction and object selection} 
\label{sec:selection}
\begin{figure}
    \centering
    \includegraphics[width=16cm]{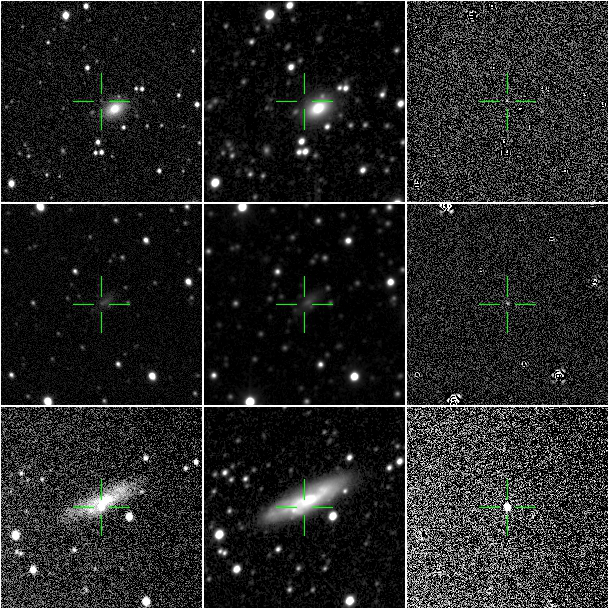}
    \caption{Three examples of triplet images, target (left), reference (middle) and difference (right) of objects that have pased the automated filters, are associated with galaxies with known redshifts and are presented to human scanners for final selection. They are confirmed supernovae SN2019shb, SN2019tua and SN2019vsa (respectively from top to bottom), the top one has a tensorflow real-bogus factor RB=0.71 and the other two have RB=1.}
    \label{fig:triplets}
\end{figure}

\begin{figure}
    \centering
    \includegraphics[width=16cm]{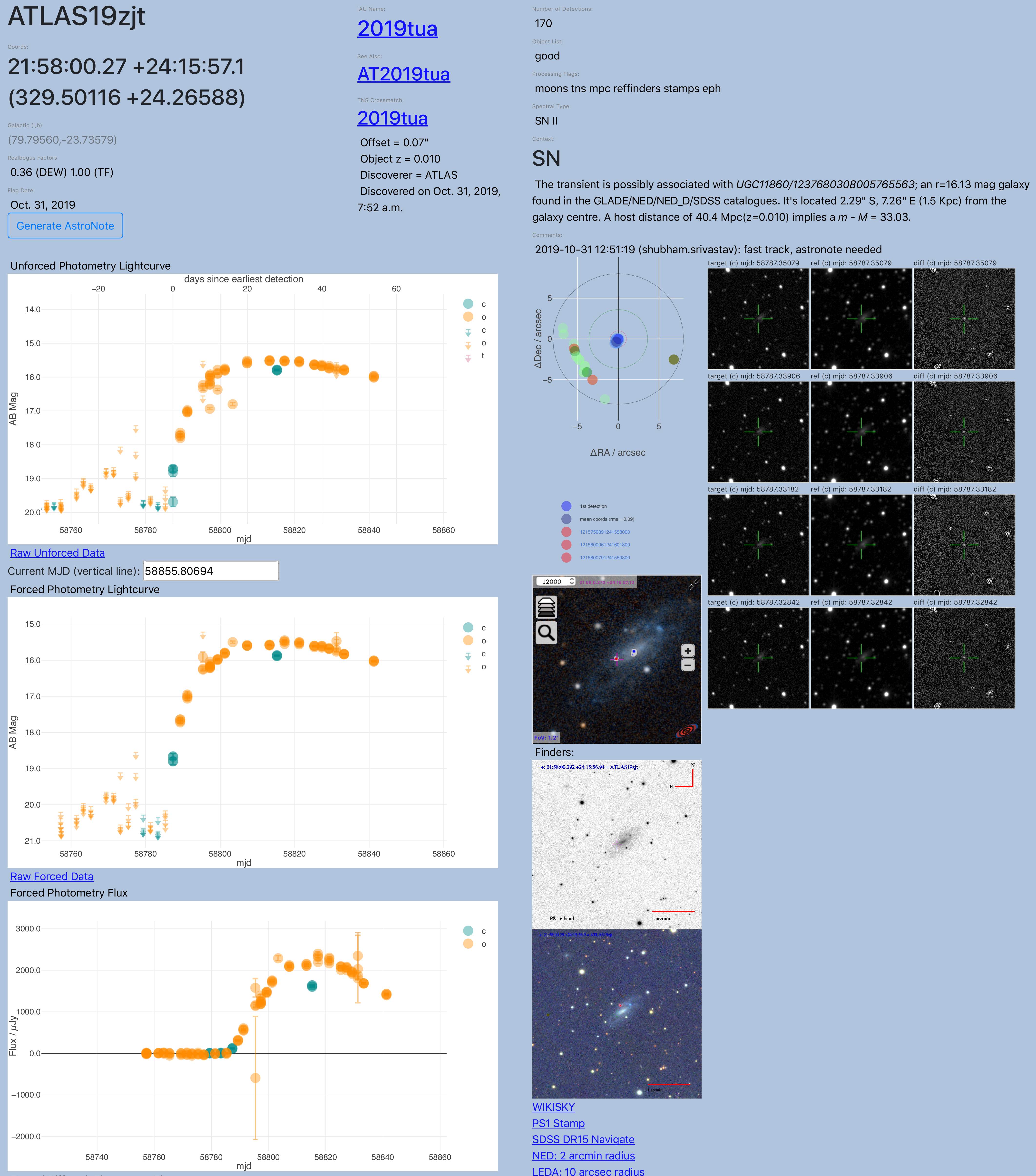}
    \caption{Candidate object page showing unforced, forced and forced photometry in flux space ($\mu$Jansky units) lightcurves.  (Non-detections in the magnitude plots are shown as arrows, which display the limiting magnitude (5$\sigma$ for unforced, 3$\sigma$ for forced.)  The most recent four rows of images on the day it was flagged are also displayed, along with a scatter plot of detections (which also shows a passing asteroid).  An Aladin-Lite widget is shown displaying Pan-STARRS imagery, and underneath this are shown two finders from Pan-STARRS data in both PS1 g-band and colour (PS1 gri). Text information on the host galaxy is shown. A crossmatch with the Transient Name Server is also displayed, and the object spectral type is shown to be a type II supernova.
    }
    \label{fig:webcandidatepage2019tua}
\end{figure}

The user is presented (via the web interface) with a triplet of postage stamps: the target image (typically 4 per night), the reference image used as the template for subtraction, and the difference image.  The scanners inspect 
these and use a web form to decide whether or not they are indeed real or bogus transients
that have a high machine learning score. 
Figure\,\ref{fig:triplets} shows examples of the triplet of images presented to the users.
As well as these triplet images, the user is presented with a range of other visualisation 
plots and tools which include the forced photometry, Aladin-lite 
\citep{2000A&AS..143...33B,2014ASPC..485..277B}
plugin showing the Pan-STARRS1
$3\pi$ archive image (if above $\delta > -30^{\circ}$, otherwise it defaults to the DSS), Galactic
coordinates, a summary of the \texttt{Sherlock} cross-matching results (along with host 
galaxy information).
Figure\,\ref{fig:webcandidatepage2019tua} shows examples of the panels displayed to users. 
At this point, human scanners decide whether to log the object in the ATLAS ``Good Objects'' list, reserve judgement and save it to a holding list (called ``Possible Objects'' ) to wait for further data, 
or to  archive it. The object remains in the database even if archived, and can be retrieved. The Transient Name Server (TNS) python API\footnote{https://wis-tns.weizmann.ac.il/content/tns-getting-started} is used to submit objects promoted to the TNS as soon as objects have been indentified as ``Good'' by the scanners.
All future detections from ATLAS that are spatially associated with an object get added to the
lightcurve, and forced photometry runs on all objects in the ``Good Objects'' and ``Possible Objects'' 
list once a day, and the lightcurve plots are updated. 
A summary of nightly ingests are given in Table\,\ref{tab:ingestnumbers}. 

{\em GW Event Tagging:}
Software has been written (in python) to download the latest LIGO-Virgo GW event Healpix maps and calculate which ATLAS objects lie within the 90\% contour region of a GW event.  Objects are tagged with the event ID, event MJD and contour value (in 10\% divisions) if they are within -10 to +21 days of the GW event.  E.g. ``Possible GW Events Association: GW190425 (MJD 58598.3458914 - 60\% contour).''  The ATLAS web interface has an additional filter facility that can be used to search for named GW events and display lists of objects that lie within the 90\% region.  This facility can be used to prioritise the scanning of relevant objects after a GW event.
This was used to discover the afterglow of GRB 170105A which was in the skymap of the 
binary black hole merger GW170104 \citep{2017ApJ...850..149S}. 

\begin{figure}
    \centering
    \includegraphics[width=16cm]{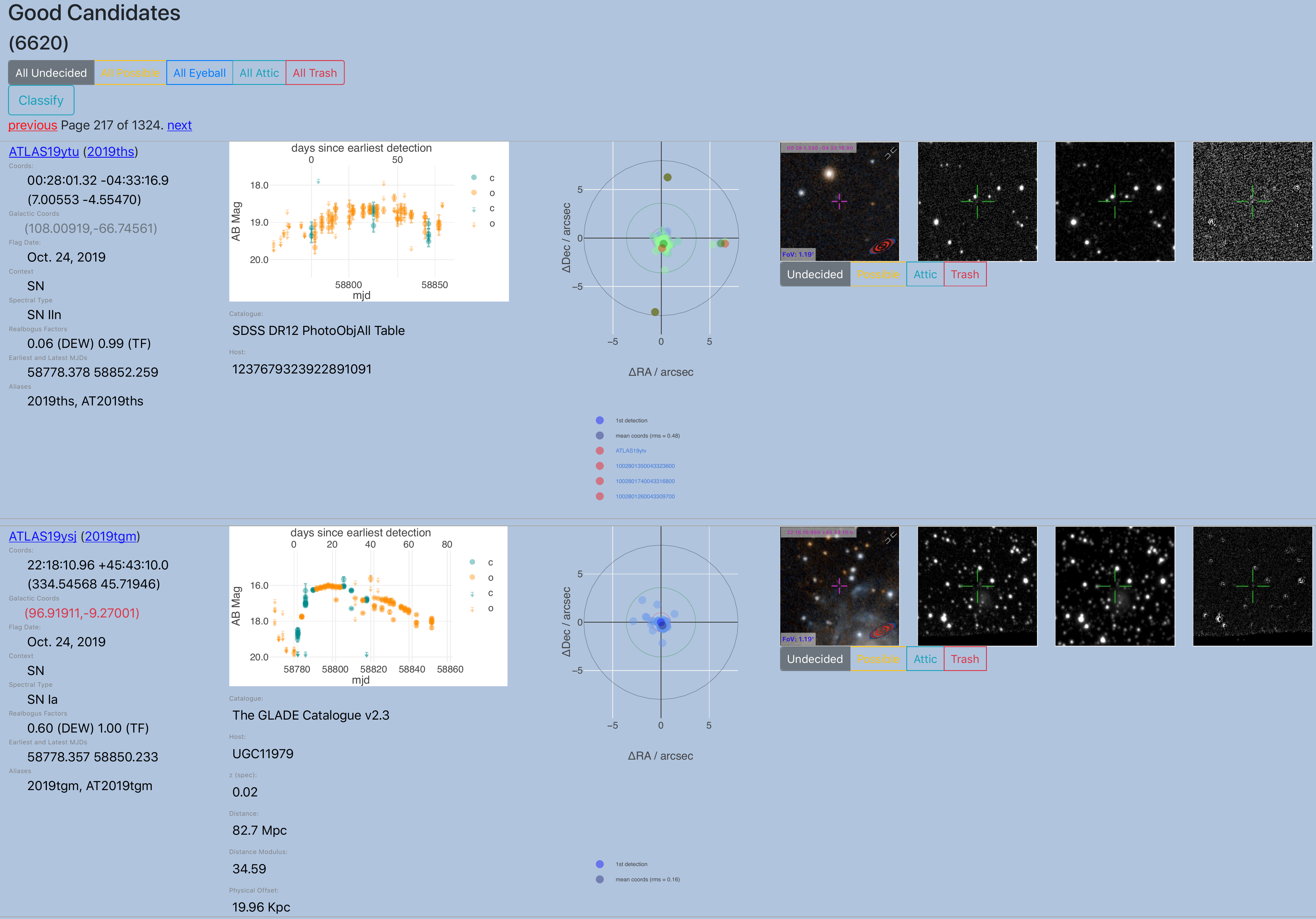}
    \caption{A quickview page showing multiple objects in the ``good'' list.  Summary information including lightcurves, scatter plots and Aladin-Lite widgets are presented, along with a representative triplet of postage stamps. Objects can be moved one-at-a time using buttons under the stamps, or all at once using ``select all'' buttons at the head and foot of the page.  Lists of objects can be filtered by (e.g.) context type, or GW event ID, and ordered by (e.g.) RB factor.  Note that the galactic coordinates are highlighted in red if the object is within galactic latitude $|b| < 15$.}
    \label{fig:webquickview}
\end{figure}

\subsection{Hardware}
\label{sec:hardware}

The processing of data is done by two rack servers -  a 28 core server with 128GB RAM for the ATLAS data processing and another identical but separate machine to host the database.  The servers are linux based (CentOS7) and are equipped with 26TB (RAID6) of spinning disk (for postage stamp storage) and 16TB of NVME SSD in a RAID0 configuration (for database storage, to reduce I/O latency). The database SSDs are backed up to spinning disk twice a week, so loss of an SSD, while destructive to the array, would not result in permanent loss of data.

ATLAS postage stamps are stored on a 26TB RAID6 spinning disk array, local to one of the above servers, while the full sized reduced ATLAS images (input, reference and difference) are stored on a bank of 4 RAID6 70TB dedicated network attached storage arrays.

The webserver requires relatively little CPU or memory, so is a separate virtual machine (running on shared university hardware) with an allocation of 4 cores and 8GB RAM. Local webserver data is stored on a small SSD with postage stamp images accessed via NFS from the above 26TB array.

Raw ATLAS data (currently about 700TB) is backed up directly from Hawaii to the QUB petabyte scale HPC cluster. Periodic ATLAS database dumps and postage stamps are also stored there.

\subsection{Software}
\label{sec:software}
Most of the software is written in Python, with \texttt{C++} (ingester) and javascript (dynamic lightcurve rendering) components.  The ingester, post ingest cutting, context classification and external survey crossmatching are marshalled by \texttt{bash} scripts called via CRON.

{\em Ingester:}
The ingester is written in \texttt{C++} (with the template functions generated by python code) and dependent on the \texttt{C++} HTM library.  The code is spawned using a Python Multiprocessing wrapper.

{\em Post Ingest Cutting and Crossmatching:}
The post ingest processing, context classification (\emph{Sherlock}) and external crossmatching code (including minor planet rejection) is written in python, with extensive utilisation of Astropy \citep{astropy:2013,astropy:2018}. Most of the code is called via the Python Multiprocessing module to make maximum use of our CPUs.

{\em Database:}
The database technology is MySQL.  The ``back-end'' storage engine chosen for most tables is the non-transactional MyISAM engine.  Although this is becoming obsolete in future, with users strongly being advised to move to InnoDB (transactional storage), the MyISAM engine is still the preferred solution for ATLAS for the time being.  Its main advantage is that the database files are a third of the size of InnoDB database files.  Given that the database is currently 6TB in size, we would require 18TB of SSD storage to run the InnoDB version just to stand still.  Another advantage is that the database files can be copied directly from one machine to another without having to apply transactional logs (assuming the source and destination machines are running the same version of MySQL).  Additionally, the way the database is currently designed (with primary keys of objects represented by their location in the sky) does not lend itself well to the InnoDB model of database row insertion (though this would be relatively easy, albeit time consuming, to rectify by the creation of a new running counter primary key).  One final advantage is MyISAM tables can be ordered by a specified index.  One property of HTMs is that spatially nearby triangles are physically close to each other in the database (because they are numerically close).  Hence if a table is ordered by the HTM index, this reduces lookup time.  The main disadvantage of MyISAM is that the whole table is locked when a single row is updated (forcing many parallel updates to be serialised internally).  However, parallel inserts are allowed, and this property is used by the ingester.

The ATLAS database, which has been running since 2015 December and contains 20 billion detections (5.7 billion objects) currently occupies about 6.2 TB (as of 2020 March 12).  The size will likely double with the introduction of the southern telescopes in Chile and South Africa.  The ATLAS postage stamps currently occupy 12TB (mostly junk or asteroid data, which is kept for future machine learning), but extraction of the stamps also requires download of the original input, reference and difference images - which currently accounts for 0.2 TB per day per telescope.  Hence a single telescope database is growing by 0.75TB per year, stamps are growing by 1.5TB per year and image data (if download continues at the current rate) will require 60TB per year per telescope.  Several times this space for backups (twice weekly and monthly) on spinning disk is also reserved. 

An additional unsolved problem is the database storage of the photometric time series images on all the individual reduced images (not the difference images) to 
provide variable star and AGN lightcurves. \cite{2018AJ....156..241H} presented the first catalogue of variable stars from ATLAS, which is 4.7 million stars, selected 
from 142 million stars that each had 100 photometric points or more. A more ambitious project would be to 
create an ATLAS billion star database which would store 1000 points on the lightcurve per year for a billion stars.  The size of this database would potentially be at least 100TB per year, assuming just 100 bytes per detection and the database would need to ingest 64\,000 detections per second, assuming 50\% utilisation.  The current ATLAS Transient Science Server infrastructure is not yet designed to cope with a database of this size and a different architecture and design would be required to serve all 
variables. 

{\em Web Interface:}
A web interface (written in python using the Django Web Framework\footnote{https://www.djangoproject.com}, Bootstrap\footnote{https://getbootstrap.com} and Plotly\footnote{https://plot.ly/javascript/})
has been built atop the database to allow the human scanners to visually inspect objects that have a reliable machine learning score.  The web interface is designed to be ``responsive'' hence will adjust how pages are rendered based on whether they are being viewed via a mobile or desktop browser.  Individual objects are displayed to the user utilising as much of the space as possible (in desktop orientation) to display the maximum amount of information without having to excessively scroll down (Figure\,\ref{fig:webcandidatepage2019tua}).
Users can be presented with lists of objects in either simple tabular format or with full rendered lightcurves, scatter plots, Aladin-Lite context, summary information and the most recent image triplet (Figure\,\ref{fig:webquickview}).  The latter allows a user to see groups of 50 (by default) objects at a time (adjustable), allowing them to archive multiple objects or retain for promotion with a single click.  The lists may be sorted by (e.g.) RB factor, latest magnitude and/or filtered to display only subsets of objects (e.g. objects tagged as SN or NT, or objects associated with a specific LIGO/Virgo event).

\section{On-sky performance and monitoring}
\label{sec:onsky}
As with any survey facility, the sensitivity of an ATLAS image is dependent on atmospheric
transparency, sky background (moon phase) and the image quality or
PSF as measured on the detector.  We monitor all of these parameters with 
measurements made on every on sky science, reduced science frame. The values are 
are written into the reduced image FITS header to allow on-sky performance monitoring and 
visualisation of long term trends.  (Most of this information is also propagated to the \texttt{.ddc} files and hence stored in the database.)  Plots of the temporal variation are given in
Figure\,\ref{fig:monitor_new_kws} for the full ATLAS history. 
Points to note on these plots are:

\begin{figure}
    \centering
    \includegraphics[width=6.6cm]{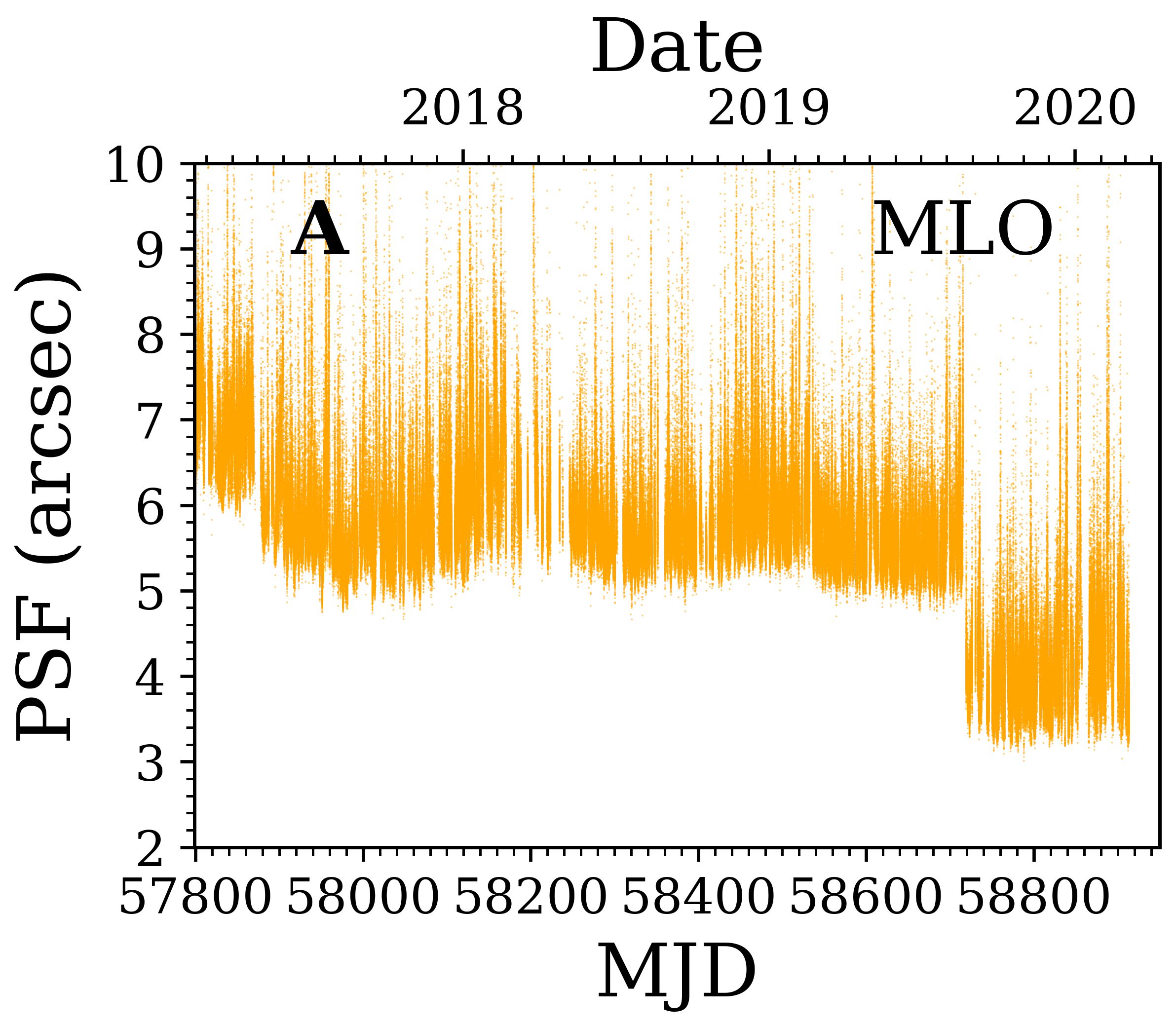}
    \includegraphics[width=8.9cm]{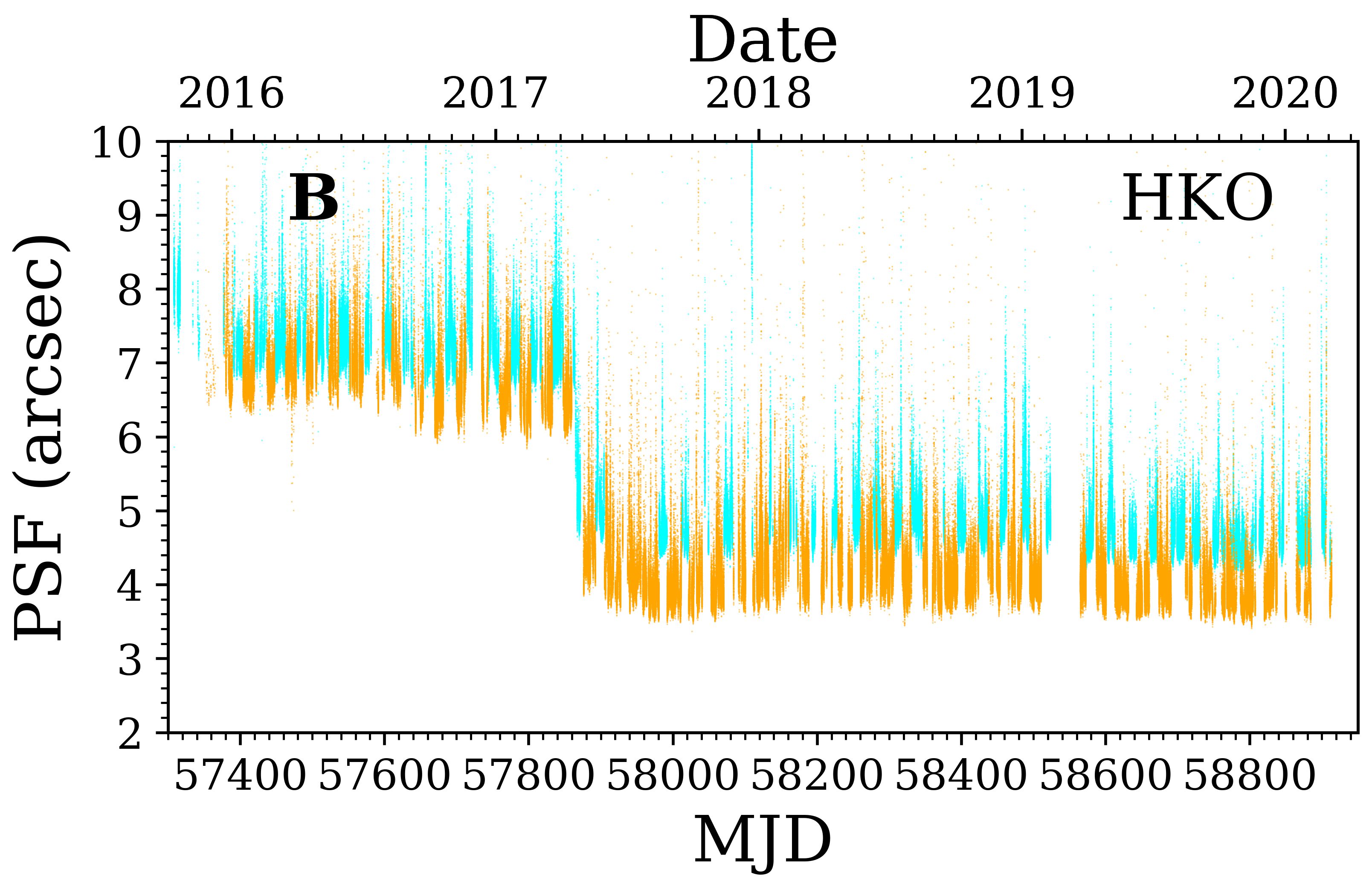}
    \includegraphics[width=6.6cm]{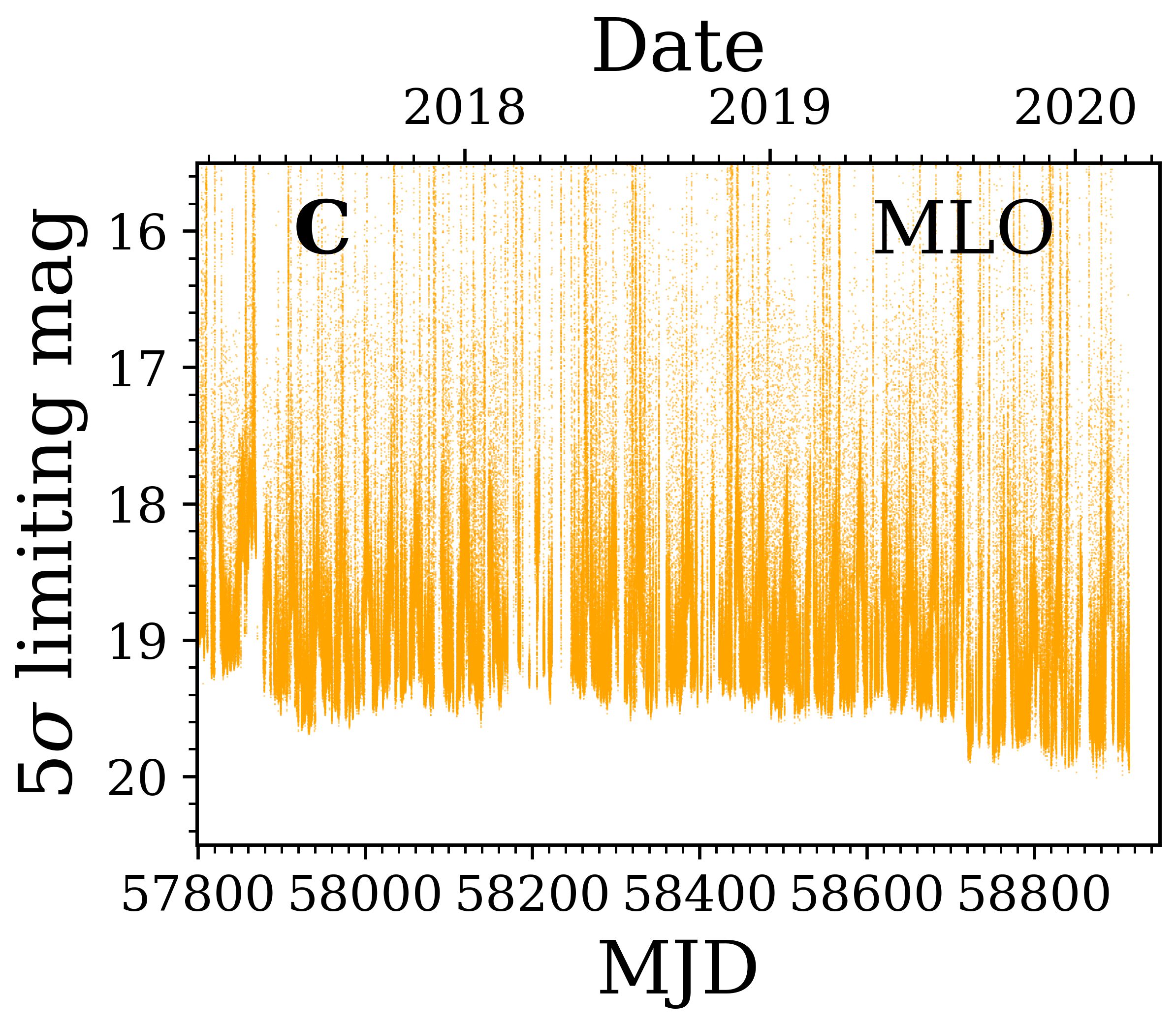}
    \includegraphics[width=8.9cm]{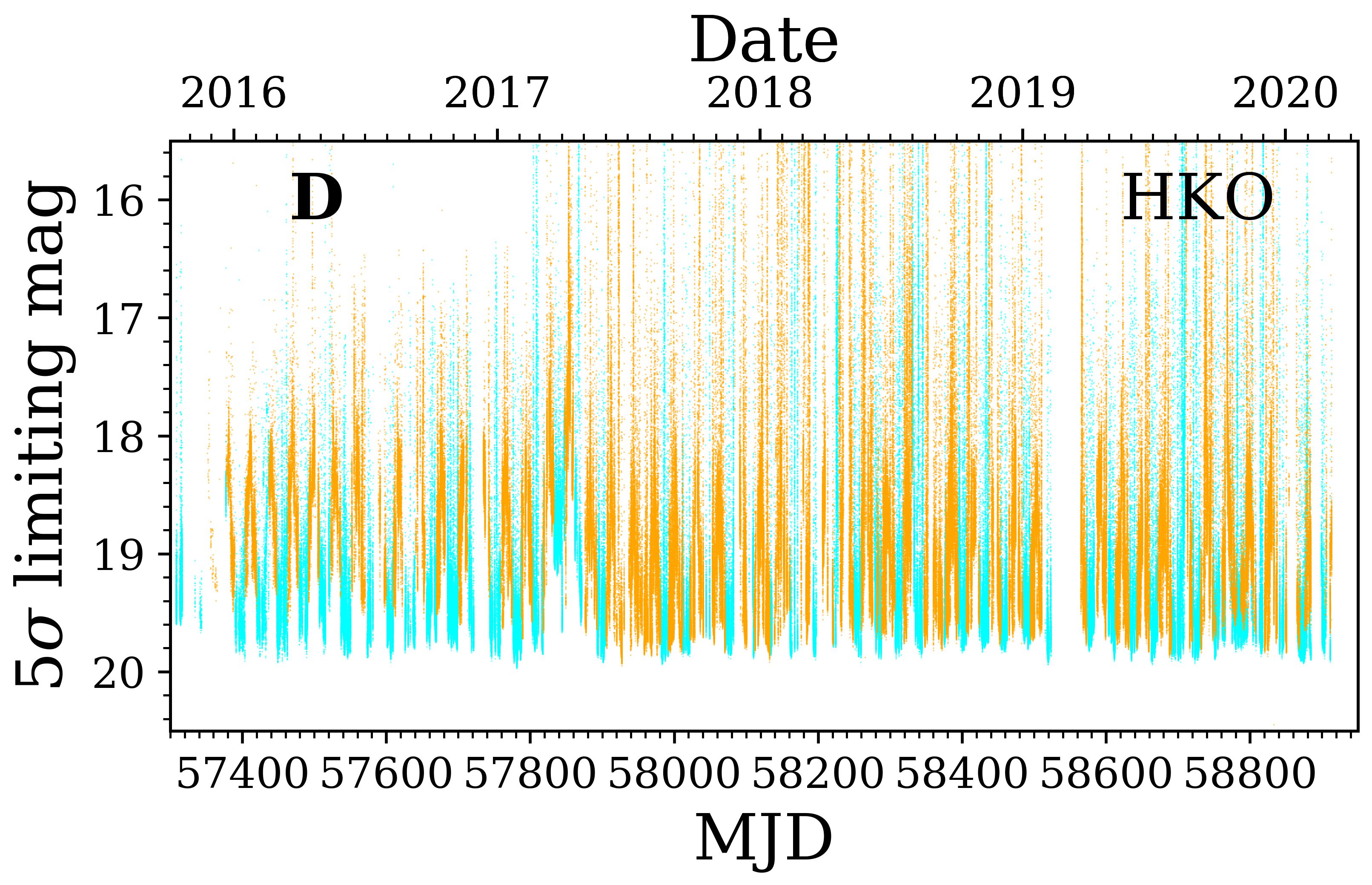}

\caption{Image quality monitoring for ATLAS HKO and MLO for their full history.  Cyan points represent measurements in the $c$ filter and orange in the $o$ filter. Recall that Mauna Loa only has had an $o$ filter in place to the present date. Note that the plots show limits \emph{per exposure}. 
{\bf A \& B:} PSF measurements - the recorded FWHM (in arcseconds) measured on the detector.
{\bf C \& D:}  The 5$\sigma$ limiting magnitude, which is a function of zeropoint variation (extinction), PSF variation and sky brightness.}
\label{fig:monitor_new_kws}
\end{figure}

\begin{enumerate}
\item The significant improvement in PSF around MJD 57864 (2017 April 21) 
is due to the replacement of the Schmidt correctors on both telescopes. The improvement is initially less obvious on ATLAS-MLO for the reasons stated below.
\item The throughput at ATLAS-HKO was observed to roll off by 0.2 mag 
between 58420 (2018 October 29) and 58605 (2019 May 02). This was due to a faulty dehumidifier which emitted an oily 
residue that gathered on the CCD window. The window was cleaned on 58606 and immediately the throughput returned to normal. The faulty dehumidifier has been replaced. 
\item The large gap at MJD=58525 (2019 February 11) to 58565 (2019 March 23) was
due to an ice storm on Haleakala, which enforced a long period of dome closure. The storm lasted just a few days but disabled electrical service to the Haleakala summit for 42 days.  The gap is also visible in Figure\,\ref{fig:Ingestion}.
\item The ATLAS-MLO sensitivity saw a slow, gradual improvement (due to improving the PSF through focus stability) 
until MJD=58715 (2019 August 20). Until that point, the
PSF measured on the ATLAS-MLO detector was significantly
worse than on ATLAS-HKO (5.5 arcsec median vs 4.0 arcsec median on HKO). 
A new STA 1600 detector was installed on MJD=58715 which immediately improved the stellar image widths to a median of 3.8 arcsec, even better than ATLAS-HKO. This was due to charge diffusion on the old CCD which degraded the FWHM (caused by a manufacturing defect). This effect corresponded to a blurring of approximately 2 pixels FWHM, in quadrature. 
\item The $5\sigma$ limiting magnitude displays the expected modulation due to lunar phase and 
the variation in sky background. 
\end{enumerate}

\begin{figure}
    \centering
    \includegraphics[width=5.3cm]{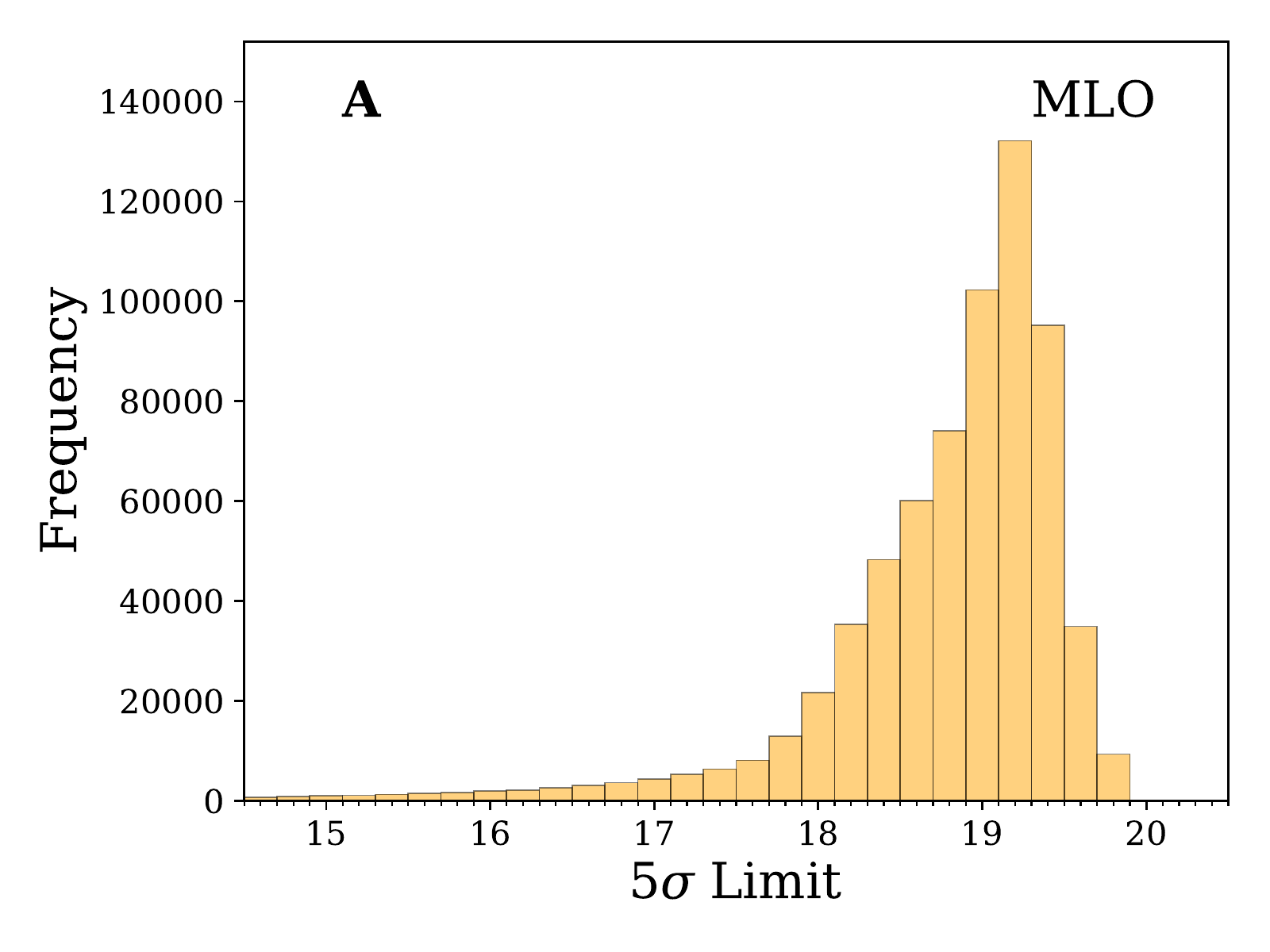}
    \includegraphics[width=5.3cm]{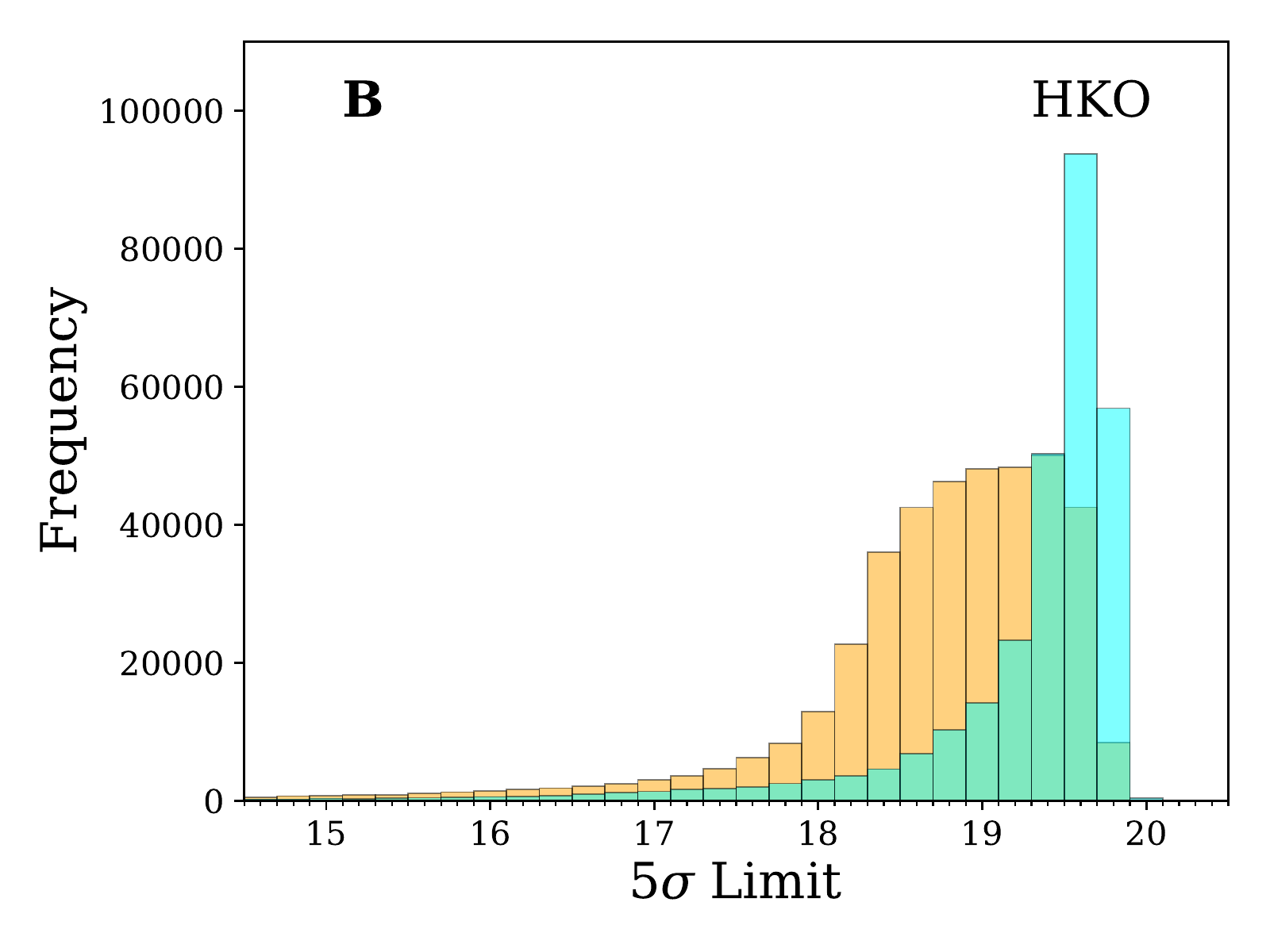}
    \includegraphics[width=5.3cm]{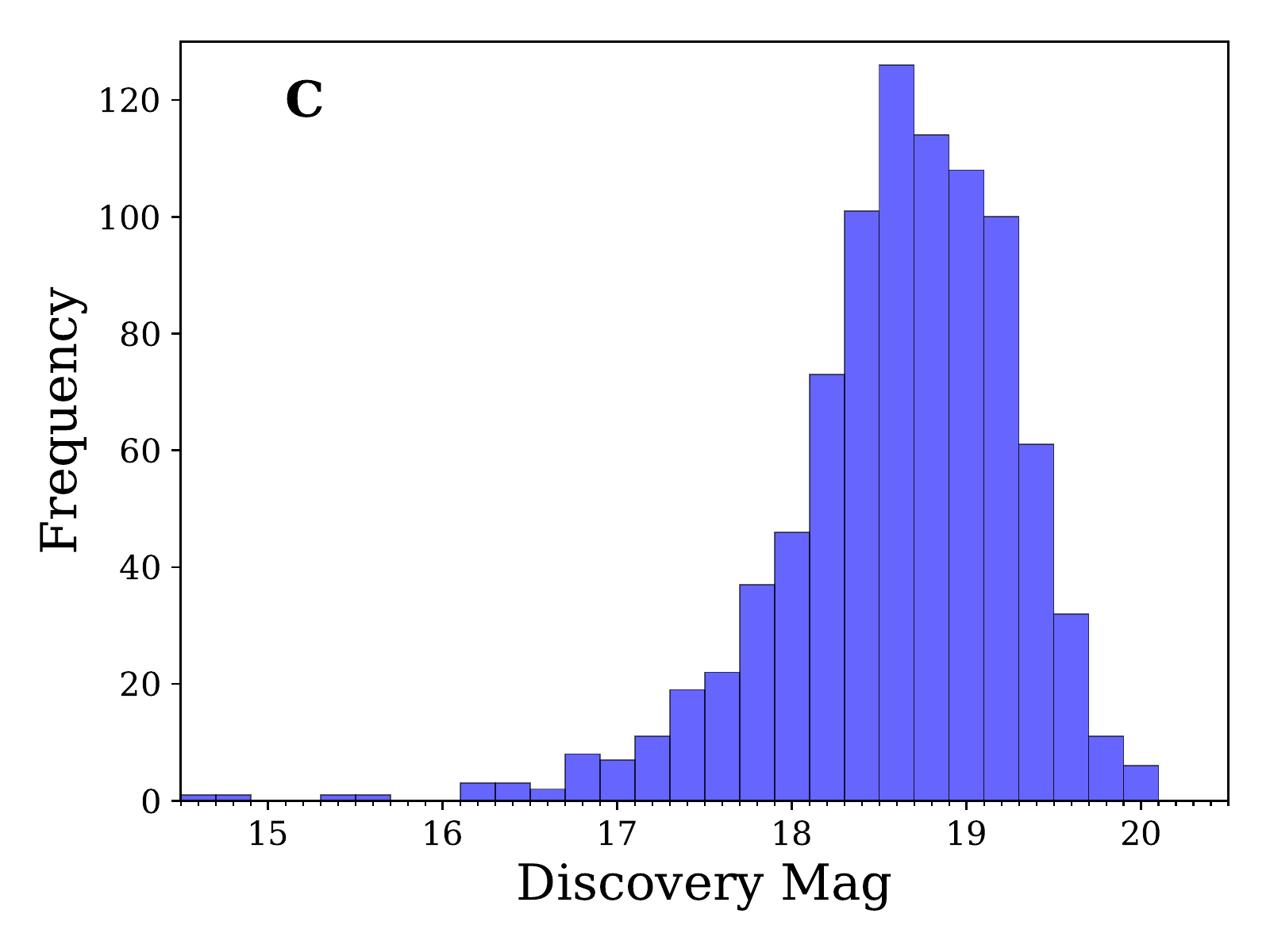}
\caption{{\bf A:} Histogram of 5$\sigma$ limiting magnitudes for MLO in the $o$ band, with the same data as in Figure\,\ref{fig:monitor_new_kws}. 
{\bf B:} Same as in A, but for HKO with the $c$ and $o$ bands plotted together. {\bf C:} The discovery magnitude distribution for all ATLAS discovered transients reported to the TNS that have been classified as supernovae.}
  \label{fig:5sigHist}
\end{figure}

The histogram of  the 5$\sigma$ limiting magnitudes for each telescope
are given in Figure\,\ref{fig:5sigHist} for the calendar year 2019. This includes all sky conditions and 
all atmospheric transparencies. As can be seen in both Figure\,\ref{fig:5sigHist} and 
Figure\,\ref{fig:monitor_new_kws} there is a long tail toward poor quality images with limiting magnitudes 
$m_{5\sigma} < 16$, which we deem not to be of science quality. These typically make up 2\% of the total science images. The median limiting magnitudes for each 
of the systems (from 2019) is 
$o < 19.0$ mag (ATLAS-HKO)
$c < 19.6$ mag (ATLAS-HKO)
and 
$o < 19.0$ mag (ATLAS-MLO). Although ATLAS-MLO has historically had poorer PSF measurements, and a similar zeropoint (i.e. similar throughout), the fact we employ it in both dark and bright time compensates since ATLAS-HKO is preferentially used when the moon is above the horizon. Since the new CCD was installed on ATLAS-MLO on MJD=58715, the median has improved to $o < 19.3$ mag. 

\begin{figure}
    \centering
    \includegraphics[width=16cm]{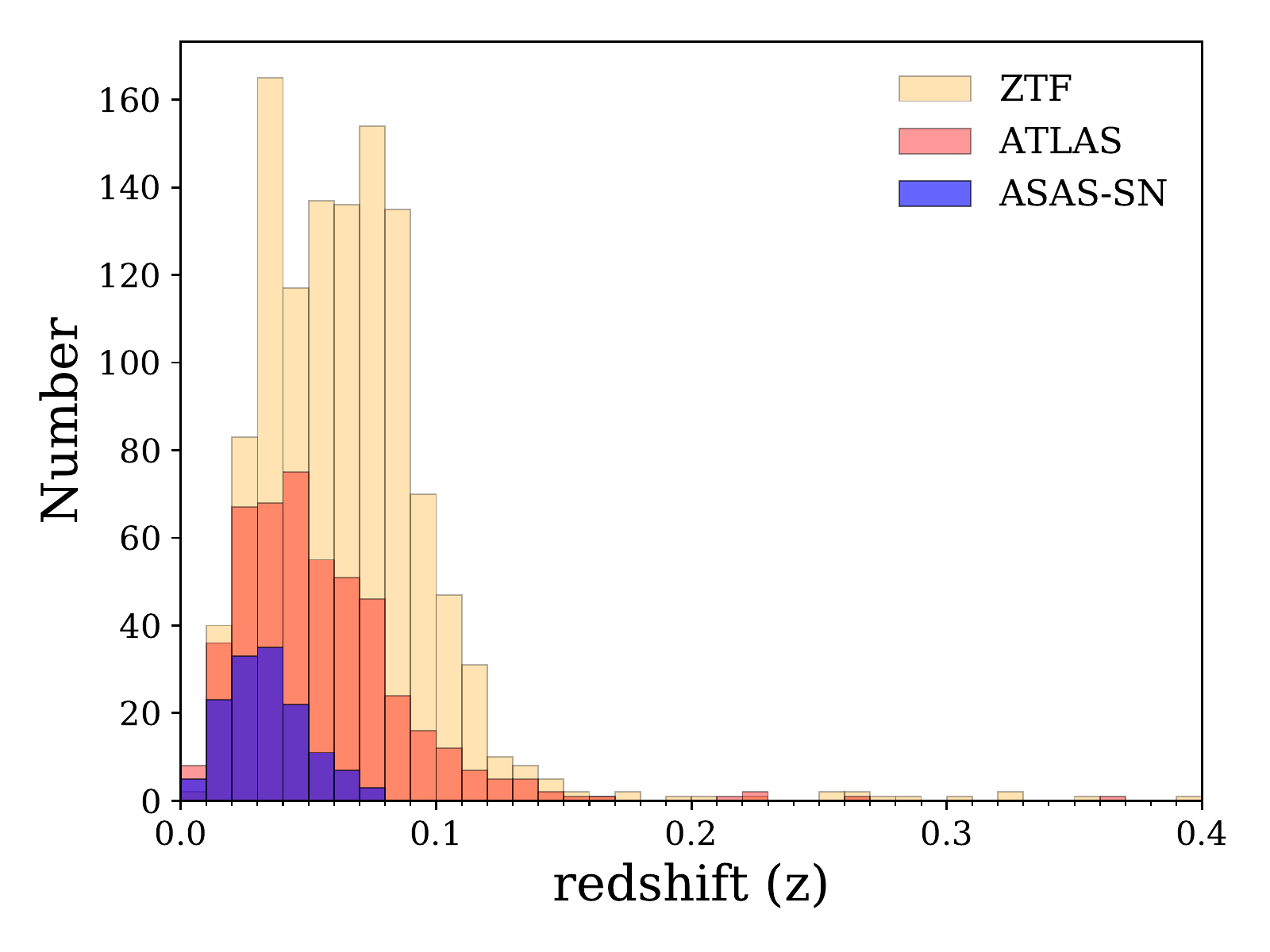}
\caption{Redshift distribution of all ATLAS discovered and detected extragalactic transients reported to the TNS in 2019 that have 
a spectroscopic redshift. For comparison, the redshift distribution of spectroscopically confirmed supernovae reported to TNS in 2019 by ZTF and ASAS-SN are also shown. It should be noted that ZTF have dedicated spectroscopic followup facilities.}
  \label{fig:zdist}
\end{figure}

\begin{figure}
    \centering
    \includegraphics[width=16cm]{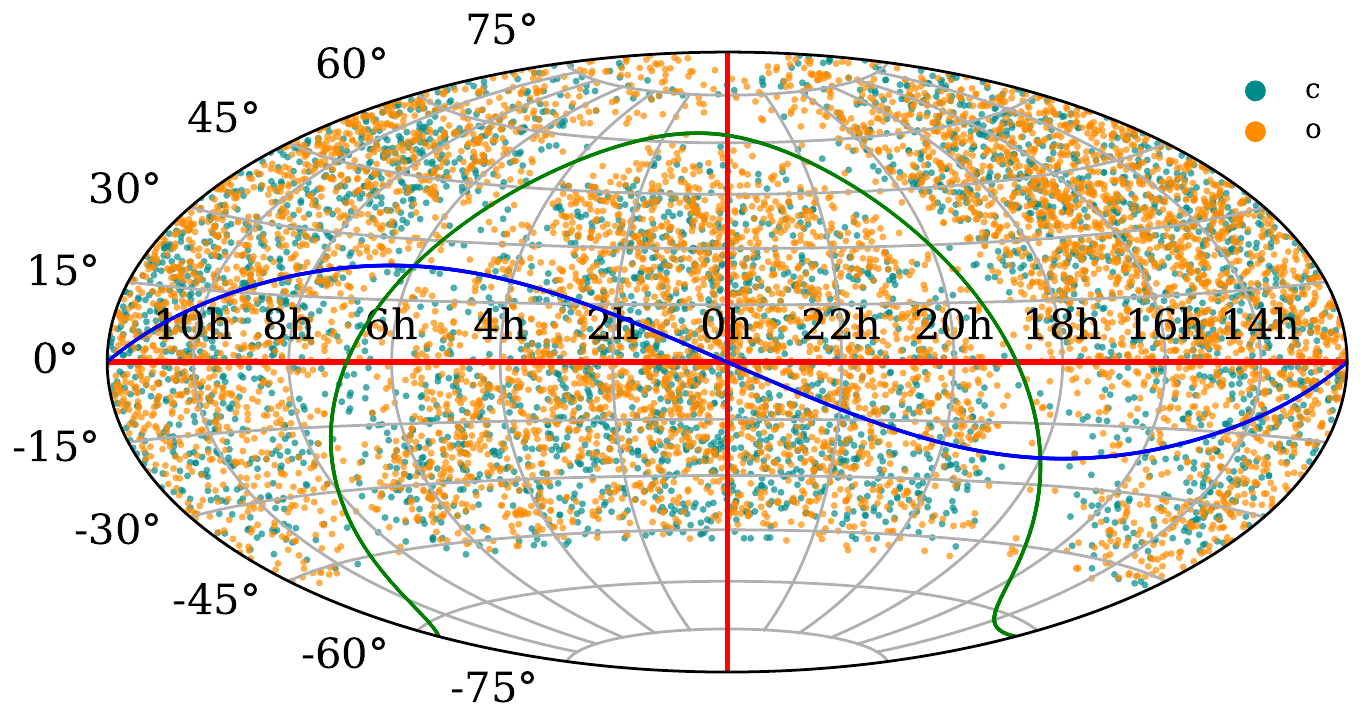}
\caption{Hammer projection of the ATLAS transients discovered from the start of the survey until 2020 March 15. Cyan and orange points represent the discovery filter of the object. Green and blue lines are the galactic plane and ecliptic respectively.}   
  \label{fig:atlasgoodhammer}
\end{figure}

\section{Summary of results and discoveries}
\label{sec:sci}
As described in Section\,\ref{sec:selection}, real astrophysical transient sources are selected by the ATLAS Transient 
Server system, with the final step being human scanning of all the candidate sources which result from
the computational processing steps. Table\,\ref{tab:ingestnumbers} indicates that 
we typically  typically have 300-400 objects per day to present for human vetting 
which results in a typical mean discovery rate or 10 to 15 real 
extragalactic transients per day. All these real sources are automatically registered on the International Astronomical Union's Transient Name Server (IAU TNS)\footnote{https://wis-tns.weizmann.ac.il/}, and a catalogue of our discoveries is easily recovered from the TNS web interface 
or through simple scripts \citep[e.g.][]{2020TNSAN...1....1K}. 

To date (2020 March 15), we have registered 5282 transients as primary discoveries, and have detected 10\,332 in total in the ``Good'' list in our  Transient Science Server database (the difference being those discovered by other surveys, mostly iPTF, ZTF, Gaia, ASASSN and Pan-STARRS). 
We aim to register extragalactic transients that are not known AGN on the TNS. Hence, as far as possible we remove 
objects that are coincident with known AGN and CVs (see Section\,\ref{sec:trans-server}) and fast declining transients 
(more than 0.5 mag within the 1\,hr quad) that are generally associated with faint red objects and are hence M-dwarf flares
\citep{2019arXiv191205549R}.
While we save these in our database as real objects, our intention is not to let these slip into the TNS reporting stream. 

We also do not report objects that are almost certain CVs, which means a very fast rise of several magnitudes
within 1-3 days (typically a gradient of $grad < -0.5\,$mag\,day$^{-1}$) from last non-detection and 
which have no obvious host object visible to $g$, $r$ or $i \simeq23$ mag
in the Pan-STARRS 3$\pi$ archive imaging. While this behaviour is almost exclusively a sign of CVs and they 
dominate the sky density of very fast rising transients, it is important not to be over zealous with this filter, since
a number of rare, fast, blue extra-galactic transients inhabit this parameter space.  In addition, 
if extragalactic transients are more than a projected distance of $R_{\rm p} \simeq 50$\,kpc from 
their host, then association with the correct host and redshift becomes confusion limited. 
The discovery of AT2018cow showed how a fast blue transient with a similar rise as a CV
could turn out to be a luminous and unusual transient at 60\,Mpc, which radiated from the 
hard x-ray to the radio
\citep{2018ATel11727....1S,2018ApJ...865L...3P,2019ApJ...872...18M}.

The selection of transients from detections on the subtractions (in the \texttt{.ddc} files) to final registration on the 
TNS depends on several steps, each of which are not guaranteed 100\% completeness. This is also true of 
other surveys, so a straight comparison of discoveries does
not yield a meaningful result. 
As a test of completeness of our processing, we extracted  transients discovered by two 
other surveys during the 6 month period 58620 (2019 May 17) to 58787 (2019 November 16), after the HKO ice-storm and the ATLAS-HKO CCD window fogging issue. 

The ASAS-SN project \citep{2019MNRAS.484.1899H} registered 88 objects as discoveries on the TNS
above $\delta = -50^{\circ}$, and we searched for these in our data. 
13 of these were discovered at RAs with large hour angles, and in an area of sky 
that ATLAS had either stopped observing for the 
season or had yet to start. We could not confirm the existence of 12 
of the 88 (and they are as yet unconfirmed on the TNS), suggesting they are not 
real. This left 
53 transients that should be in the ATLAS database and 
registered as objects. All were detected on the ATLAS 
difference images (multiple times at $S/N>5$), and  
all 53 objects were in the ATLAS Transient database. 
Three objects did not
make the initial filter of $\geq$3, good quality
detections, at greater than 5$\sigma$, on any one night and two had low machine 
learning RB scores. A 4\% missed detection rate from the 
machine learning is consistent with the levels we set for selection from the ROC curve discussed in 
Section\,\ref{sec:selection}. 
One further object was rejected as
being associated with a compact galaxy
labelled as stellar in the PS1 catalogue (therefore
Sherlock classified it as a VS). Therefore our computer 
filtering
from end to end appears susceptible to a 10\% incompleteness. 
A further 8 objects were not promoted as real sources
by human scanners, as they judged them not to be reliable
astrophysical transients on the first night that they were detected. 
This indicates the level of uncertainty 
introduced when scanners have a small number
of detections on which to judge real or bogus (typically
introducing a 15\% hit on efficiency). This human scanning incompleteness 
usually arises from low significance detections on the first night of
detection, and the objects are archived, while subsequent nights 
will produce much more reliable and secure detections. (See improvements in section \ref{sec:conc} item \ref{conclusions:objectresurrection}.) A search for ZTF 
objects with a peak $ r < 19$ mag and at least 4 detections in \texttt{Lasair}, and 
a cross-correlation with ATLAS objects showed similar results. The end to 
end comparison showed we had approximately 80\% efficiency with the losses
split roughly equally between automated computer filtering (basic cuts and machine 
learning) and human decisions. 

\begin{figure}
    \centering
    \includegraphics[width=1.0\linewidth]{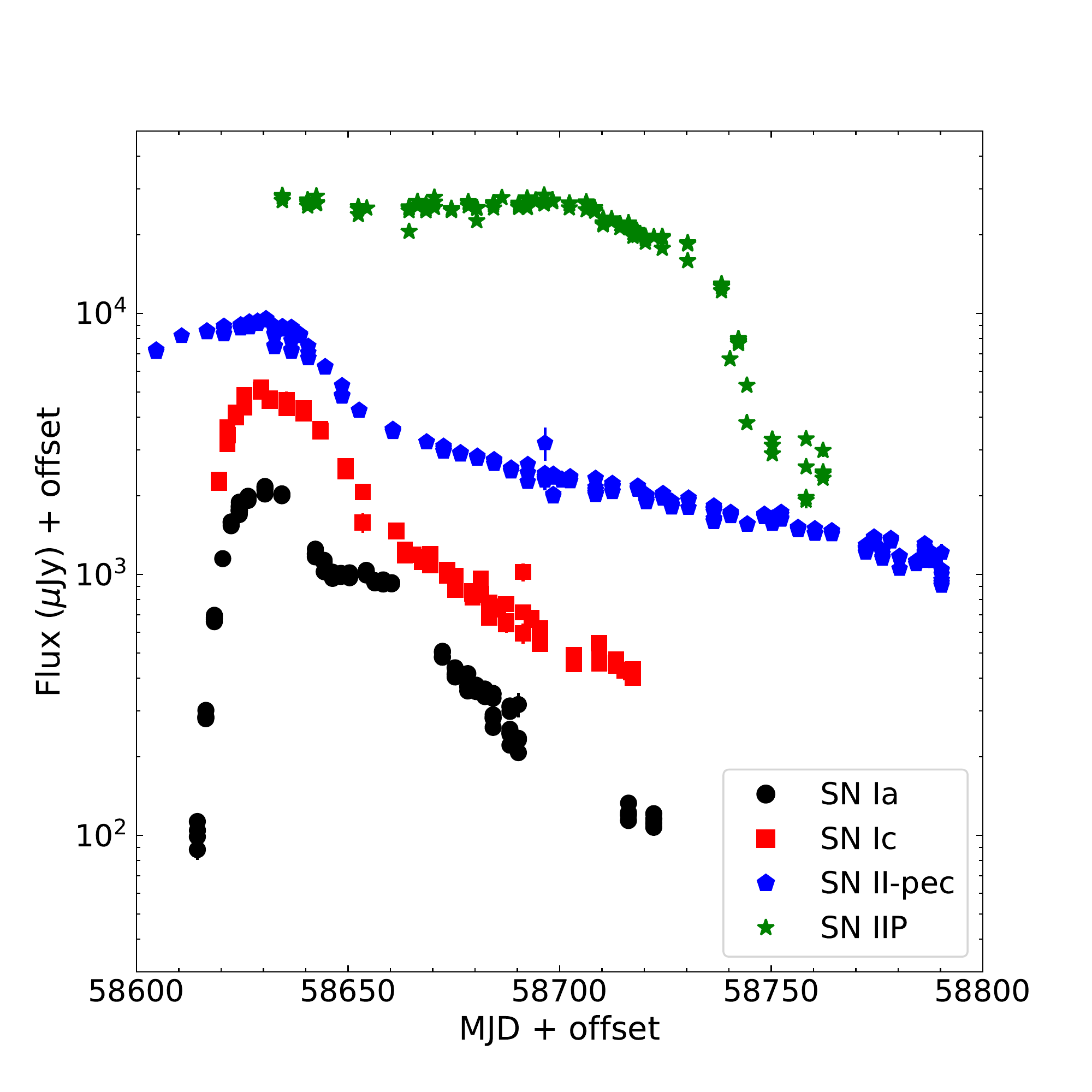}
    \caption{Well-sampled $o$-band ATLAS light curves of 4 nearby SNe of types Ia, Ic, IIP and II-peculiar.
    The ``shoulder'' or secondary peak in the light curve of the SN Ia $\sim 30$ days after the primary maximum is seen clearly, whereas the stripped envelope SNe (SESNe: Ib, Ic and IIb) do not exhibit this feature. The light curves were shifted along the x and y axes for clarity.} 
    \label{fig:lightcurves}
\end{figure}

In Figure\,\ref{fig:zdist} we illustrate the redshift distribution of the spectroscopically classified 
ATLAS transients in 2019. We include only the objects that were formally discovered by ATLAS (which we define
as first report to the TNS) and exclude those detected by ATLAS but were discovered by other surveys. (Spectroscopically confirmed objects reported to TNS by ZTF and ASAS-SN in 2019 are also shown for comparison.) 
At a redshift of $z = 0.15$, a discovery magnitude of $m \sim 19.5$ corresponds to $M \sim -19.6$ mag, or a typical Ia at peak. Those SNe beyond  $z = 0.15$ are significantly brighter than the 
bulk of the normal population. Of the 15 SNe found at $z>0.15$, 7 are super-luminous SNe (SLSNe), whereas the remainder include bright 
Ia (Ia 91T-like or Ia-CSM), IIn and Ic broad-line events.
Figure\,\ref{fig:lightcurves} shows example light curves of different classes, indicating
that in the $o$-filter, the near-infrared secondary peak in normal type Ia SNe is clearly 
detected, allowing SNe Ia to be distinguished from SNe Ib/c. 
This will assist SN rate calculations for the supernovae for which we do not have
a spectroscopic classification.

\begin{figure}
    \centering
    \includegraphics[width=0.45\linewidth]{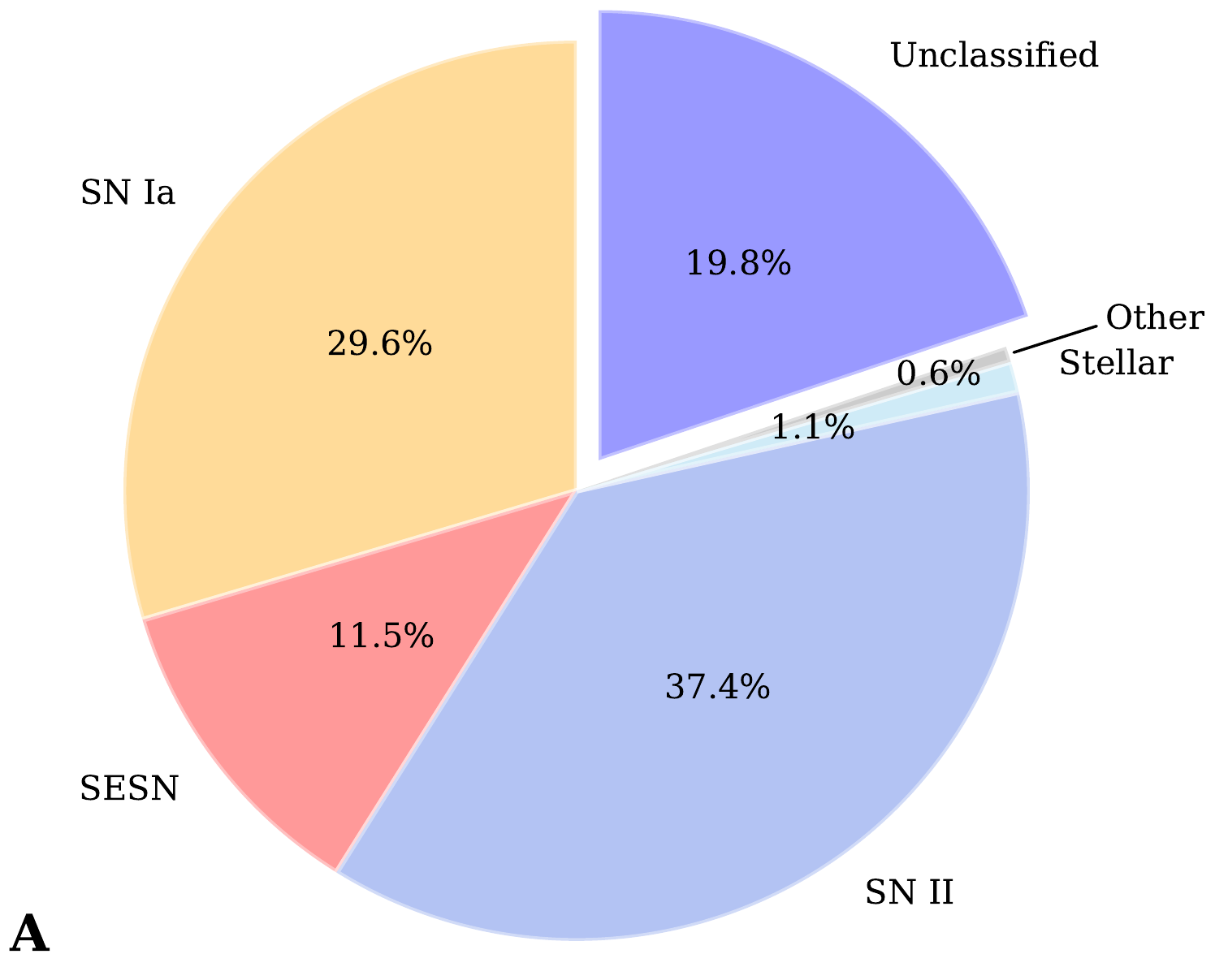}
        \includegraphics[width=0.45\linewidth]{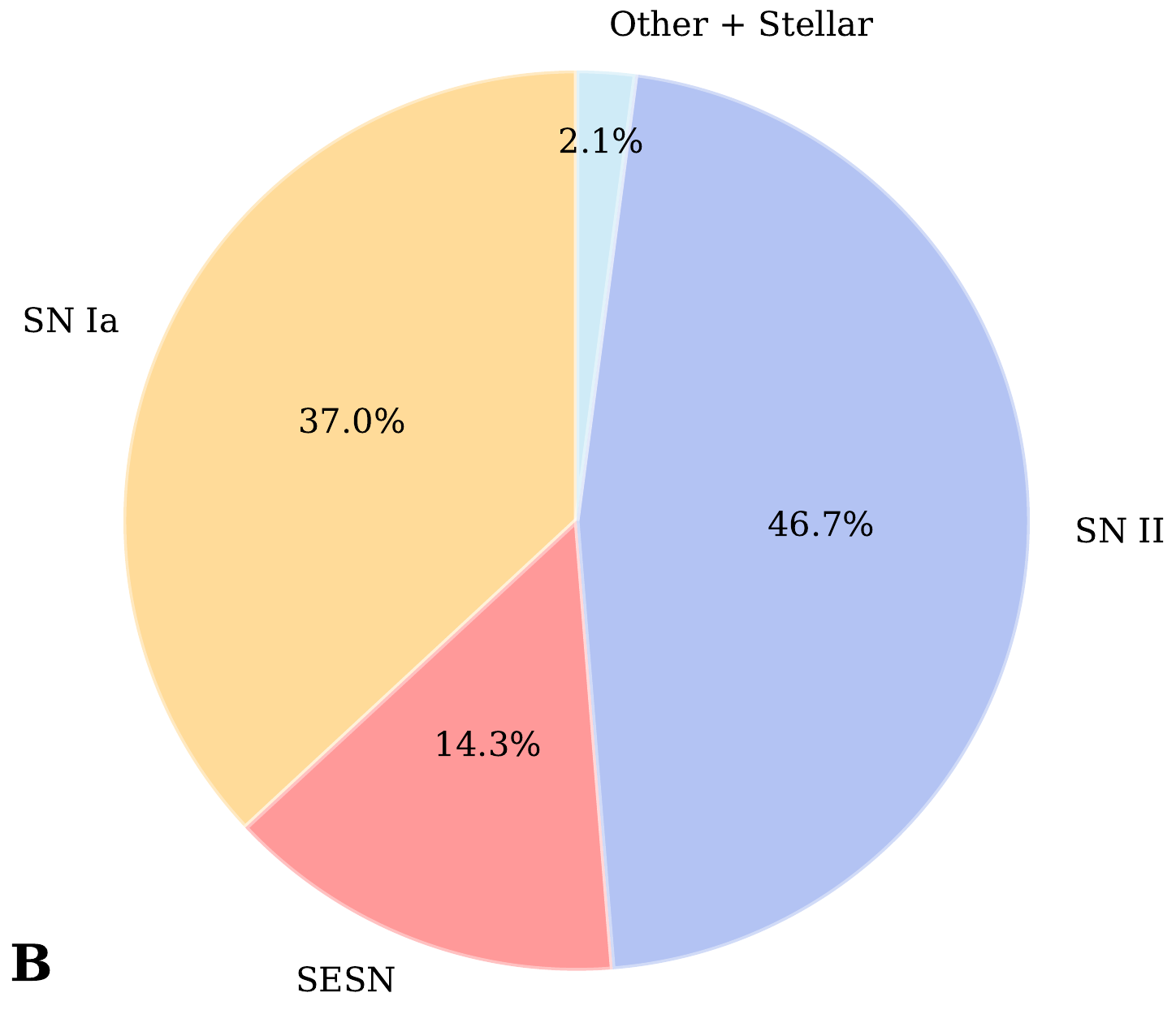}
    \caption{Pie charts of the spectroscopic classifications of ATLAS detected transients associated with galaxies with distances within 100\,Mpc. {\bf A:} All ATLAS transients.  {\bf B:} Subset of spectroscopically classified transients. SESN represents the family of stripped envelope SNe that includes types Ib/c, IIb and Ic broad-line. Stellar transients include Galactic CVs and variable stars as well as extragalactic novae, whereas the ``Other'' category includes a TDE, ILRTs and LBVs. A total of 540 transients were included, out of which 107 transients are without a spectroscopic classification. 
    }   
    \label{fig:SNtypes_all}
\end{figure}

\subsection{Spectroscopic types within the local 100\,Mpc volume} 
An interesting scientific survey that 
ATLAS will contribute to is measuring complete, volumetric rates of transients
within the Local Universe. At a distance of 
100\,Mpc ($\mu=35$), we are sensitive to 
transients of absolute magnitude of 
$M \lesssim -16$ (in regions of low Milky Way extinction) and is an upper limit to the 
distance within which we will carry out 
this survey. It is also approximately 
the upper distance limit to which we
would be sensitive to faint, fast fading
kilonova type objects like AT2017gfo
\citep[which peaked at $M_{r,i} = -16$ mag, within the first 24\,hrs; see the early combined $r$ and $i$ photometric evolution in ][]{Andreoni2017,Arcavi2017,Coulter2017,SoaresSantos2017,Smartt2017,Valenti2017}. 

A total number of 540 transients, associated with host galaxies within 100 Mpc, were detected by ATLAS and reported to the TNS between 2017 September 22 (MJD 58018.5) and 2019 December 03 (MJD 58820.2).  These galaxies had either distances or spectroscopic redshifts (in our three primary catalogues: NED, GLADE and SDSS) and the association radius
was set at a sky projected distance of 
50\,kpc in  \emph{Sherlock}.   These include transients discovered by other surveys, but all were 
independently detected by ATLAS. 
Out of these, 107 transients (19.8\%) do not have a reported classification on TNS. Of the 433 transients with a reported spectroscopic classification, 6 (1.1\%) were stellar sources, i.e. foreground Galactic CVs and variable stars or extragalactic novae. Tidal disruption events (TDEs), intermediate luminosity red transients (ILRTs) and luminous blue variable outbursts (LBVs) together constitute a small fraction (0.6\%) of the classified transients. The breakdown of the classifications is shown in 
Figure\,\ref{fig:SNtypes_all} and we emphasise that this should not be taken as a true measure of relative rates since 
we have carried out no correction for completeness as a function of absolute magnitude of each type. It is purely 
illustrative of the numbers of transients detected which are associated with galaxies within this distance limit and 
shows that about 1 object per day is found which is plausibly within 100\,Mpc. As an illustrative comparison we can compare our 
results to the Lick Observatory Supernova Search sample \citep[LOSS;][]{2011MNRAS.412.1441L,2011MNRAS.412.1473L}, which was a fully corrected, volume limited survey of catalogued galaxies within 60 - 80\,Mpc, from a sample of 180 SNe in total.  
The relative rate for core collapse SNe (CCSNe) in our sample (433 objects after removing unclassified transients) is 61.0\%, whereas SNe Ia account for 37.0\%. In comparison, the relative rate of CCSNe and SNe Ia was 60.6\% and 37.7\% respectively, in the Lick Observatory Supernova Search sample \citep[LOSS;][]{2011MNRAS.412.1473L}.

Within this 100\,Mpc volume, 
we can estimate the simple  ``observed'' volumetric rates for supernovae. 
Assuming that ATLAS covers the whole sky down to $\delta \sim -50^{\circ}$ (88\% of 4$\pi$) and 
that $\pm60$\,degrees in RA from the sun is unobservable and lost (33\% of 4$\pi$), then ATLAS covers approximately 60\% of the 
full $4\pi$ sky with a 2 day cadence. The number of type Ia and CCSNe that we have detected 
within 100\,Mpc implies that the lower limit to the rate of supernova is 
$R_{Ia} = 5.3 \times 10^4$ Gpc$^{-3}$ yr$^{-1}$ for SNe Ia and 
$R_{CC} = 9.0 \times 10^4$ Gpc$^{-3}$ yr$^{-1}$ for CCSNe. 
This is based on all ATLAS detections, not just on discoveries and 
these volumetric rates represent a hard lower limit, since correction factors involving sensitivity and sky coverage have not been taken into account. When we calculate such corrections they will
only have the effect of significantly increasing the rates since we 
are not 100\% efficient in detecting SNe within this volume.  The above is based on 
assuming the 540 transients contain 2\% of non-supernovae (which we know is the fraction
of the 433 classified transients), and that the rest are supernovae with a relative
rate the same as in the 433 classified events. 
One may worry about a bias introduced by the selection of transients for 
spectroscopic follow-up. False positives such as mistaken CVs, 
outbursting stars or non-astrophysical artefacts would lie in this un-classified list
simply because those with spectroscopic resources considered them not likely to be 
real supernovae. We have visually inspected all lightcurves of these unclassified 
107 objects. There are 6 which are likely either CVs, stars or poor subtractions
leading to false positives. All the other 101 have clear supernova like lightcurves 
and are associated with 100\,Mpc galaxies with no clear candidate for a background
galaxy observed to Pan-STARRS1 3$\pi$ depths. Many of the objects are toward the
fainter end of the distribution, $m\gtrsim18.5$ mag and some have incomplete lightcurves
due to seasonal and or weather constraints. However they are all very likely 
SNe of some type and a future paper will address classification of these from their
lightcurves, with a particular focus on the  ``shoulder" feature in the $o$-band 
lightcurve which distinguishes type Ia SNe from the others. The 6 possible 
impostor objects do not influence the above calculation.

It is interesting to note that 
 these observed rates, with no corrections are already higher than those previously published at low redshift.  For  type Ia SNe, \citet{2011MNRAS.412.1473L} report a rate of $3\pm0.6\times 10^4$\,Gpc$^{-3}$ yr$^{-1}$ within 80\,Mpc 
 for targeted galaxy searches and \cite{2019MNRAS.486.2308F} report 
 $\approx 2 - 3 \times 10^4$ Gpc$^{-3}$ yr$^{-1}$ within $ z \lesssim 0.09$ (400\,Mpc). For CCSNe, LOSS find a rate of 
$7.1\pm1.6 \times 10^4$ Gpc$^{-3}$ yr$^{-1}$ \citet{2011MNRAS.412.1473L}, which is higher (but not seriously discrepant, considering the uncertainties) to that from 
\cite{1999A&A...351..459C} of $4\pm2 \times 10^4$ Gpc$^{-3}$ yr$^{-1}$.  Our estimate of
the type Ia rate is nearly a factor 2 higher than currently accepted, which is quite
a difference and the discrepancy would only increase when a proper efficiency and survey 
simulation calculation is carried out. The CCSN rate is less discrepant, at 30\% higher than
the LOSS \citet{2011MNRAS.412.1473L} value. However the completeness
correction for CCNSe is likely to be much larger than for type Ia, due to their fainter 
luminosities \citep{2011MNRAS.412.1441L,2011MNRAS.412.1473L}, and we are certainly far from 
complete for a $M_o\simeq -16$ mag core-collapse SN at 100\,Mpc (since, $o = \mu + M_o + A_o \gtrsim 19$ mag).
Therefore it is  possible that 
some of the SNe which are classified as type Ia on the TNS are actually type Ic (or Ib), 
since this mis-identification is often made \citep{Clocchiatti2000}. This would decrease the Ia rate and increase
the CCSN rate, although the total SN rate would be preserved, and that is already 50\% higher
than the published values. The ATLAS SN counts give a total SN rate of 
$1.4\times10^{5}$ \,yr$^{-1}$\,Gpc$^{-1}$, compared to the LOSS total of $1.0\times10^{5}$ \,yr$^{-1}$\,Gpc$^{-1}$. 
Another possibility is that both the type Ia and CCSN rate really are significantly
higher than measured in LOSS and indeed all other low redshift surveys. This could be 
 due to a higher specific SN rate  in lower mass galaxies, 
meaning targeted galaxy surveys will be biased against the high rates of production in 
low mass galaxies. This type of effect has been suggested by ASAS-SN for type Ia SNe
 \citep[see the ASASSN results in an all-sky survey][]{2019MNRAS.484.3785B}. 
If the core-collapse SN rate is significantly higher than the LOSS rates, and
if such a systematic underestimate were also present at higher redshifts, then 
it may close the discrepancy between the cosmic core-collapse rate and the
star formation rate from other indicates as noted by 
\cite{2011ApJ...738..154H}. Although \cite{2012A&A...537A.132B} find that very local SN rates and 
the traditional star-formation rate indicators (H$\alpha$ and FUV fluxes) 
are discrepant in the opposite sense.  To address these important problems, 
we will undertake further careful calculation of the local SN rates
in future papers. At face value, these simple estimates indicate that the accepted local 
Universe supernova rates could be too low by a factor 1.5 to 2.


\section{Conclusions and future improvements }
\label{sec:conc}

The ATLAS system now routinely surveys the whole sky visible from Hawaii (above 
$\delta > -50^{\circ}$) every two nights with the twin telescope system. This results
in 10-15 extragalactic transients discovered or detected (where the latter means discovered first by other surveys) every night with a 5$\sigma$ detection limit of $o < 19$ mag. 
All discoveries are made public promptly, however the spectroscopic 
classification completeness is only about 25\%. The processing at QUB, during 
day time while the ATLAS telescopes are observing produces rapid discoveries 
now routinely during the Hawaiian night. Computer algorithms for filtering 
detections focus on real-bogus classifications, removal of moving objects and stars, 
and a combination of boosted decision trees and machine learning provides an 
efficient method to present users with a short list of real extragalactic 
transients. However human vetting is still essential in this last step to 
ensure an optimal balance between false positive rate and missed detection rate. 
The ATLAS discoveries not only enable rapid follow-up of interesting 
sources \citep[e.g. AT2018cow][]{2018ApJ...865L...3P} but will provide complete
statistical samples within the local volume of 100\,Mpc. 

Looking forward, we foresee the following improvements to the system
\begin{enumerate}

    \item Addition of two more units in the southern hemisphere at the South African Astronomical Observatory, and El Sauce in Chile. These will allow the full sky to be covered every night with our quad sequence and transients to be discovered routinely within 12-24\,hrs. An assessment of the data flow speed to QUB and what extra hardware is required is essential. It will be prudent to review if the QUB Transient Science Server should continue to be the main processing node or if Chilean/South African resources (such as the ALeRCE architecture) can improve the speed and reliability and if there should be multiple sources of data release for the transients.   
    \item Process the co-added $4\times30$\,sec exposures each night as one, deeper 120\,sec exposure. We now make a co-added nightly stack of the difference images, which results in a single image deeper by 0.75 magnitudes. Implementing our 5$\sigma$ detection process on these
    images should enable us to reach $o < 20$ mag and $c < 20.3$ mag routinely. We are experimenting with the real-bogus balance and how to robustly select objects with one detection on nightly stack. During image subtraction, it would be worthwhile to pursue improvements to the convolution calculation, optimised for our sampling and point-spread-function. This should help
    decrease the false positive rate on the co-added difference images. 
        \item \label{conclusions:objectresurrection} Development of code that will resurrect any object that has been rejected by human based on low signal-to-noise detections if new detections arise 
        that are more significant. 
    \item Application of a machine learning algorithms that will classify objects primarily by their lightcurve
  \citep[e.g. RAPID][]{2019PASP..131k8002M}. However combining this with  
    host object information (galaxy redshift, offset from the galaxy centre, morphology, and colour) should lead to improvements in the prediction of supernova/transient type from an incomplete lightcurve. 
    \item Rapid release of transients as Kafka alerts, as is currently done with ZTF, and combining automatically with ZTF to provide extra time resolution. 
    \item Improvements to the input galaxy catalogue, with more spectroscopic redshifts and any reliable photometric redshift measurements
    \citep[e.g.][]{Takada2014,DESI2016,2019Msngr.175....3D}.
    \item Further, closer ties to follow-up programmes to increase spectroscopic classification rates and early data. 
    \item Creation of a public user interface that provides access to all ATLAS lightcurve data and all information that is currently proprietary to the science teams. This is a significant effort to provide both the computing and software resources to support a large, public user base. 
    \item Creation of a large database of variability of every object in the sky brigther than $o \sim 19$ mag. And to provide public access as in the previous point. 
\end{enumerate}

\section{Acknowledgements}
 This work has made use of data from the Asteroid Terrestrial-impact
  Last Alert System (ATLAS) project. ATLAS is primarily funded to search
  for near earth asteroids through NASA grants NN12AR55G, 80NSSC18K0284,
  and 80NSSC18K1575 (under the guidance of Lindley Johnson and Kelly Fast); 
  byproducts of the NEO search include images and
  catalogs from the survey area.  The ATLAS science products have been
  made possible through the contributions of the University of Hawaii
  Institute for Astronomy, the Queen's University Belfast, the Space
  Telescope Science Institute, and the South African Astronomical Observatory.
  We acknowledge support for transient science exploitation from the EU FP7/2007-2013 ERC Grant agreement n$^{\rm o}$ [291222], STFC Grants ST/P000312/1, ST/N002520/1, ST/S006109/1 and support from the QUB Kelvin HPC cluster, and the QUB International Engagement Fund. TWC acknowledges EU Funding under Marie Sk\l{}odowska-Curie grant agreement No 842471.

\bibliographystyle{aa}
\bibliography{libsjs,libdry,libkws}


\end{document}